\documentclass[11pt]{article}
\pdfoutput=1
\usepackage{graphicx}

\usepackage{jheppub}

\usepackage{xcolor}

\edef\restoreparindent{\parindent=\the\parindent\relax}
\usepackage{parskip}
\restoreparindent
\usepackage{autobreak}
\usepackage{bbm}
\usepackage{braket}
\usepackage{comment}
\usepackage{mathtools}
\usepackage{tikz}
\usetikzlibrary{arrows,arrows.meta,intersections, calc,positioning,decorations.pathreplacing,decorations.pathmorphing}
\usetikzlibrary{patterns}
\usetikzlibrary{decorations.markings}
\tikzset{decorate,decoration={snake,amplitude=.4mm,segment length=2mm,post length=1mm}/.style={decorate, decoration=snake}}

\def\Tr{{\rm Tr}}
\def\d{{\rm d}}
\def\i{{\rm i}}

\def\CH{{\cal H}}

\def\d{\mathrm{d}}

\begin{document}


\title{Probing Hawking radiation through capacity of entanglement}
\author[a]{Kohki Kawabata,}
\author[b]{Tatsuma Nishioka,}
\author[a]{Yoshitaka Okuyama}
\author[c]{and Kento Watanabe}

\affiliation[a]{Department of Physics, Faculty of Science,
The University of Tokyo, \\
Bunkyo-ku, Tokyo 113-0033, Japan}
\affiliation[b]{Yukawa Institute for Theoretical Physics, Kyoto University,\\
Kitashirakawa Oiwakecho, Sakyo-ku, Kyoto 606-8502, Japan}
\affiliation[c]{Center for Quantum Mathematics and Physics (QMAP), \\
Department of Physics, University of California, Davis, CA 95616 USA}

\emailAdd{kawabata-koki382@g.ecc.u-tokyo.ac.jp}
\emailAdd{tatsuma.nishioka@yukawa.kyoto-u.ac.jp}
\emailAdd{okuyama@hep-th.phys.s.u-tokyo.ac.jp}
\emailAdd{kewatanabe@ucdavis.edu}

\abstract{
We consider the capacity of entanglement in models related with the gravitational phase transitions.
The capacity is labeled by the replica parameter which plays a similar role to the inverse temperature in thermodynamics.
In the end of the world brane model of a radiating black hole the capacity has a peak around the Page time indicating the phase transition between replica wormhole geometries of different types of topology.
Similarly, in a moving mirror model describing Hawking radiation the capacity typically shows a discontinuity when the dominant saddle switches between two phases, which can be seen as a formation of island regions.
In either case we find the capacity can be an invaluable diagnostic for a black hole evaporation process.
}

\preprint{YITP-21-08}

\arxivnumber{2102.02425}

\maketitle

\section{Introduction}
Black holes occupy a distinguished position not only as existing constitutes of our universe but also as a theoretical testing ground for searching a theory of quantum gravity that unifies general relativity and quantum mechanics.
While there is no doubt that a black hole obeys the laws of thermodynamics and owns the entropy proportional to the area of the horizon \cite{Bekenstein:1972tm,Bekenstein:1973ur}
the thermodynamic character of black holes poses a longstanding paradox when they are regarded as quantum mechanical objects that evolve unitarily in time.
Taking quantum effects into account, black holes begin to emit a thermal radiation and end up evaporating to nothing in a finite time \cite{Hawking:1974sw}.
If we throw objects there, we seemingly lose the information about the objects because a thermal black hole radiation does not depend on what has fallen, and it evaporates completely. 
Hawking raised his famous information paradox in \cite{PhysRevD.14.2460} by showing the indefinite increase of radiation entropy for an evaporating black hole, which leads to the breaking of the unitary evolution in a quantum gravity theory.
To reconcile with the unitarity, Page considered the entanglement entropy for the radiation between the inside and outside of evaporating black holes and suggested that the linear increase of the entropy caused by the radiation at early time must stop at some typical timescale and start to decrease to zero as black holes evaporate \cite{Page:1993wv}.
Reproducing the Page curve in a model of black holes with radiation is, however, still challenging and has led to considerable efforts so far (see e.g., \cite{Raju:2020smc} for an excellent review and references therein).

The idea of associating entropy with geometric quantities was cultivated by 
Ryu and Takayanagi, who proposed a holographic formula equating the entanglement entropy for a spacial region in quantum field theory to the area of a minimal surface anchored on the boundary in the dual gravitational theory \cite{Ryu:2006bv,Ryu:2006ef}.
The Ryu-Takayanagi formula is a manifestation of the generalized gravitational entropy \cite{Lewkowycz:2013nqa} which derives from a gravitational path integral using the replica method for entanglement entropy:
\begin{align}
    S_A \equiv \lim_{n\to 1}\frac{1}{1-n}\log \Tr\, \rho_A^n\ .
\end{align}
In a field theory, $\Tr\,\rho_A^n$ is given by the partition function $Z_n$ on a manifold with singular surface of deficit angle $2\pi(1-n)$ along the boundary $\partial A$ of a region $A$ at some timeslice.
In the bulk spacetime, $Z_n$ represents the partition function on a replica geometry with $n$ boundaries and the singularity along $\partial A$ extends to the bulk to form the Ryu-Takayanagi surface.
There are numerous extensions of the formula including time-dependent theories \cite{Hubeny:2007xt,Dong:2016hjy}, quantum corrections \cite{Faulkner:2013ana}, and R\'enyi entropies \cite{Dong:2016fnf}.
The most general formulation \cite{Engelhardt:2014gca} expands on the notion of the generalized second law \cite{Bekenstein:1974ax} and states that the entropy of the boundary region $A$ is given by the generalized entropy consisting of the area of a minimal quantum extremal surface (QES) and the entanglement entropy of a matter across the QES.

Recently, a proposal for computing the entropy of quantum systems entangled with gravitational systems, called the island formula was presented \cite{Almheiri:2019hni} and shown to hold in a certain class of two-dimensional dilaton gravity coupled with matter of large degrees of freedom \cite{Almheiri:2019qdq,Penington:2019kki} (see also \cite{Almheiri:2020cfm} for a review).\footnote{One can always find a local solution of replica wormholes near the QES, but the global solutions are not guaranteed to exist in general (see e.g., \cite{Hartman:2020swn} for discussion).
} 
The derivation proceeds much like the Ryu-Takayanagi formula with a gravitational path integral method but
a significant difference is there between the two formulas. The Ryu-Takanayagi formula picks up only the replica geometry with disconnected replicas while the island formula has a substantial contribution from the replica wormhole with all replicas connected in gravitational regions, which instructs us to include a region, named an island, behind a black hole horizon as a part of the entangling region in calculating the entanglement entropy of the Hawking radiation.
The island formula has been exploited to resolve the information problem by successfully reproducing the Page curve in various types of black holes \cite{almheiri2019islands,Balasubramanian:2020xqf,hartman2020islands,Anegawa:2020ezn,Hashimoto:2020cas,Hartman:2020swn,Gautason:2020tmk,Almheiri:2019psy,Alishahiha:2020qza,Ling:2020laa,Bhattacharya:2020uun}.
For further progress in replica wormholes and island formula see also e.g., \cite{Akers:2019nfi,Chen:2019uhq,Nomura:2019qps,Suzuki:2019xdq,Kusuki:2019hcg,Rozali:2019day,Balasubramanian:2020coy,Chen:2020jvn,Bak:2020enw,Agon:2020fqs,Balasubramanian:2020hfs,Krishnan:2020oun,Krishnan:2020fer,Li:2020ceg,Chandrasekaran:2020qtn,Dong:2020uxp,Geng:2020qvw,Chen:2020uac,Chen:2020hmv,Hernandez:2020nem,Akal:2020ujg,Kirklin:2020zic,Liu:2020gnp,Piroli:2020dlx,Nomura:2020ska,Pollack:2020gfa,Marolf:2020xie,Chakravarty:2020wdm,Marolf:2020rpm,Bousso:2020kmy,Stanford:2020wkf,Giddings:2020yes,Chen:2020tes,Karlsson:2020uga,Engelhardt:2020qpv,Verlinde:2020upt,Harlow:2020bee,Chen:2020ojn,Hsin:2020mfa,Akal:2020wfl,Murdia:2020iac,Goto:2020wnk,Geng:2020fxl,Caceres:2020jcn,Manu:2020tty,Miao:2020oey,Miao:2021ual}.

\begin{figure}[ht!]
	\centering
	\begin{tikzpicture}[transform shape]
 \draw[very thick,->,>=stealth] (0,0) to (6,0) node [right, font=\normalsize] {$\text{time}$};
	            \draw[very thick,->,>=stealth] (0,0) to (0,3);
 \draw[thick, orange!100] (0,0) to (2.5,2.5);
  \draw[thick, orange!100] (2.5,2.5) to (6,2.5) node [right, font=\normalsize, black] {$S_{R}$};
  
  \draw[thick, blue!100] (0,0) to[out=10,in=-120] (1.65,0.85);
  \draw[thick, blue!100] (3.35,0.7) to (6,0.7) node [right, font=\normalsize, black] {$C_{R}$};
  \draw[thick, dotted, black!80] (2.5,0.85) circle (0.85) node[font=\LARGE, black!100] {$?$};
  
	\end{tikzpicture}
	\caption{A typical shape of the entanglement entropy (orange) and an asymptotic form of the capacity of entanglement (blue) at early and late times for black holes with the Hawking radiation.
    The entropy grows linearly and saturates after the Page time while the capacity is vanishing at early time and approaches some value at late time.}
	\label{fig:capacity_asymptotic}
\end{figure}
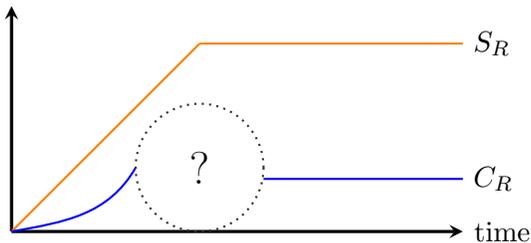

The goal of this paper is to probe the Hawking radiation process through another quantum information measure known as the capacity of entanglement:
\begin{align}\label{def_capacity}
    C_A \equiv \lim_{n \to 1} n^2 \partial_n^2 \log  \Tr\, \rho_A^{n}  \ ,
\end{align}
which is to entanglement entropy what the heat capacity is to thermal entropy when the replica parameter $n$ is seen as the inverse temperature \cite{Yao:2010woi,Nakaguchi:2016zqi}.
Some aspects of the capacity of entanglement were investigated both in holographic systems \cite{Nakaguchi:2016zqi} and in field theories \cite{Yao:2010woi,deBoer:2018mzv,Nakagawa:2017wis}, but it remains open to what extent the capacity can be a practical measure of entanglement in a broader context.
We will examine the characteristics of the capacity of entanglement in toy models of 
radiating black holes by taking $A$ to be a region $R$ outside the black holes, especially around the Page time (see figure \ref{fig:capacity_asymptotic}).
Since the contribution from the replica geometry with nontrivial topology is essential in the gravitational path integral, we expect the capacity of entanglement for the radiation is also capable of capturing the replica wormholes, or equivalently the island region as in the island formula for entanglement entropy.

At early time a fully disconnected replica wormhole dominates in the gravitational path integral.
This explains the linear growth of the entropy for the radiation region $R$ due to the Hawking radiation while the geometry typically has a vanishing contribution to the capacity (see section \ref{ss:early_late}).
On the other hand, a fully connected replica wormhole starts to compete with the fully disconnected one around the Page time and take over at late time.
Then the entropy either saturates or decreases depending on whether black holes are eternal or evaporating, resulting in a continuous Page curve.
As for the capacity we will see it is sensitive to the phase transition at the Page time, showing a crossover or discontinuity.
Note that the capacity does not equal the derivative of the entropy with respect to the radiation time, but it will turn out to play a similar role as the latter.
To our best knowledge, this is the first case where the capacity of entanglement probes a phase transition of any kind.

The structure of this paper is as follows.
In section \ref{ss:ToyModel}, we treat a quantum mechanical model of evaporating black holes known as the end of the world (EOW) brane model introduced in \cite{Penington:2019kki} that is simple enough for us to perform the exact gravitational path integral with all possible replica wormholes taken into account.
We show that the capacity of entanglement shows a peak around the Page time due to the change of the dominant saddle from the replica geometries in the model.
In section \ref{ss:MovingMirror} we then move onto two-dimensional holographic conformal field theories (CFTs) in moving mirrors which mimic radiating black holes and reproduce the Page curves \cite{Akal:2020twv}.
We find the capacity becomes discontinuous when the dominant saddle to the entropy switches between two phases, corresponding to black hole with and without an island region.
Finally we devote section \ref{ss:Discussion} to discussions and future directions.

\section{End of the world brane model}\label{ss:ToyModel}
In this section, we will consider a toy $2d$ gravity model of an evaporating black hole in the Jackiw-Teitelboim (JT) gravity \cite{Teitelboim:1983ux,Jackiw:1984je} with an end of the world (EOW) brane entangled with an auxiliary system \cite{Penington:2019kki}. 
In this simple model, island contributions to entanglement entropy in the auxiliary system are obtained from replica wormhole contributions to the R\'enyi entropies. For large dimension of the systems, the replica wormholes contributions can be calculated by summing up all planar topologies in the full gravitational path-integral as performed in \cite{Penington:2019kki}. It can be done analytically in the microcanonical ensemble and numerically in the canonical ensemble. For both cases, we will find smooth curves of the capacity.

\subsection{Early and late time behaviors}\label{ss:early_late}
First of all, we will briefly review the simple gravity model \cite{Penington:2019kki} and discuss about asymptotic behaviors of the entanglement entropy and the capacity of entanglement at early and late time (sufficiently away from the Page time), where only the topology of replica wormholes matters.
More detailed analysis of replica geometries around Page time will be deferred to the following subsections.

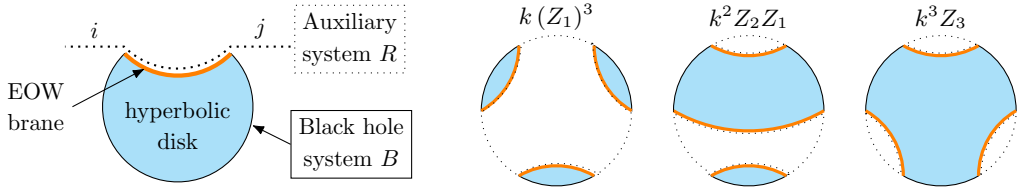
\begin{figure}
    \centering
    \begin{tikzpicture}[scale=0.5, every node/.style={scale=0.8}]
\draw[fill=cyan!30] (45:2) arc[start angle=315, end angle=225,radius=2cm] 
-- (135:2) arc[start angle=135, end angle=405,radius=2cm]--cycle;
\draw[orange, ultra thick] (45:2) arc[start angle=315, end angle=225,radius=2cm];
\draw[dotted, thick] (3,1.6) -- node[above] {$j$} (1.4,1.6) arc[start angle=315, end angle=225,radius=2cm] 
-- node[above] {$i$} (-3,1.6);
\draw (0,-1/2) node[align=center] {hyperbolic\\ disk};
\draw[{Latex[round]}-] (-0.9,1)--(-2.8,0) node[left, align=center] {EOW \\ brane};
\draw[{Latex[round]}-] (2,-0.5)--(3,-1) node[right, align=center, draw] {Black hole \\ system $B$};
\draw (3.1,1.8) node[right, align=center, draw, dotted] {Auxiliary \\ system $R$};
\end{tikzpicture}
\qquad
\begin{minipage}[t]{0.5\linewidth}
\begin{tikzpicture}[scale=0.5]
\draw[fill=cyan!30] (0:2) arc[start angle=0, end angle=60,radius=2cm] 
-- (60:2) arc[start angle=180, end angle=240,radius=2cm] --cycle;
\draw[very thick, orange] (60:2) arc[start angle=180, end angle=240,radius=2cm];
\draw[fill=cyan!30] (120:2) arc[start angle=120, end angle=180,radius=2cm] 
-- (180:2) arc[start angle=300, end angle=360,radius=2cm] --cycle;
\draw[very thick, orange] (180:2) arc[start angle=300, end angle=360,radius=2cm];
\draw[fill=cyan!30] (240:2) arc[start angle=240, end angle=300,radius=2cm] 
-- (300:2) arc[start angle=60, end angle=120,radius=2cm] --cycle;
\draw[very thick, orange] (300:2) arc[start angle=60, end angle=120,radius=2cm];
\draw[dotted] (62:2) arc[start angle=62, end angle=118,radius=2cm]
-- (118:2) arc[start angle=358, end angle=302,radius=2.2cm] 
-- (182:2) arc[start angle=182, end angle=238,radius=2cm]
-- (238:2) arc[start angle=120, end angle=60,radius=2.2cm]
-- (302:2) arc[start angle=302, end angle=358,radius=2cm]
-- (358:2) arc[start angle=238, end angle=178,radius=2.2cm]--cycle;
\draw (0,2.5) node {\footnotesize $k\, (Z_1)^3$};
\end{tikzpicture}
\quad
\begin{tikzpicture}[scale=0.5]
\draw[fill=cyan!30] (0:2) arc[start angle=0, end angle=60,radius=2cm] 
-- (60:2) arc[start angle=300, end angle=240,radius=2cm] 
-- (120:2) arc[start angle=120, end angle=180,radius=2cm] 
-- (180:2) arc[start angle=240, end angle=300,radius=4cm] --cycle;
\draw[very thick, orange] (60:2) arc[start angle=300, end angle=240,radius=2cm];
\draw[very thick, orange] (180:2) arc[start angle=240, end angle=300,radius=4cm];
\draw[fill=cyan!30] (240:2) arc[start angle=240, end angle=300,radius=2cm] 
-- (300:2) arc[start angle=60, end angle=120,radius=2cm] --cycle;
\draw[very thick, orange] (300:2) arc[start angle=60, end angle=120,radius=2cm];
\draw[dotted] (62:2) arc[start angle=62, end angle=118,radius=2cm]
-- (118:2) arc[start angle=242, end angle=298,radius=2cm]--cycle;
\draw[dotted] (180:2) arc[start angle=182, end angle=238,radius=2cm]
-- (238:2) arc[start angle=120, end angle=60,radius=2.2cm] 
-- (302:2) arc[start angle=302, end angle=358,radius=2cm]
-- (358:2) arc[start angle=300, end angle=240,radius=4cm]--cycle;
\draw (0,2.5) node {\footnotesize $k^2 Z_2 Z_1$};
\end{tikzpicture}
\quad
\begin{tikzpicture}[scale=0.5]
\draw[fill=cyan!30] (0:2) arc[start angle=0, end angle=60,radius=2cm] 
-- (60:2) arc[start angle=300, end angle=240,radius=2cm] 
-- (120:2) arc[start angle=120, end angle=180,radius=2cm] 
-- (180:2) arc[start angle=60, end angle=0,radius=2cm] 
-- (240:2) arc[start angle=240, end angle=300,radius=2cm] 
-- (300:2) arc[start angle=180, end angle=120,radius=2cm] --cycle;
\draw[very thick, orange] (60:2) arc[start angle=300, end angle=240,radius=2cm];
\draw[very thick, orange] (180:2) arc[start angle=60, end angle=0,radius=2cm];
\draw[very thick, orange] (300:2) arc[start angle=180, end angle=120,radius=2cm];
\draw[dotted] (62:2) arc[start angle=62, end angle=118,radius=2cm]
-- (118:2) arc[start angle=242, end angle=298,radius=2cm]--cycle;
\draw[dotted] (182:2) arc[start angle=182, end angle=238,radius=2cm]
-- (238:2) arc[start angle=2, end angle=58,radius=2cm]--cycle;
\draw[dotted] (302:2) arc[start angle=302, end angle=358,radius=2cm]
-- (358:2) arc[start angle=122, end angle=178,radius=2cm]--cycle;
\draw (0,2.5) node {\footnotesize $k^3 Z_3$};
\end{tikzpicture}
\end{minipage}
\caption{[Left] The geometry of $\langle \psi_i | \psi_j\rangle_B$ in the EOW brane model. The hyperbolic disk (blue) has the asymptotic boundary and terminates on the EOW brane (orange). [Right] The replica geometries for the moment \eqref{eq:nth moment} of $\rho_R$ with $n=3$. There are three ways to connect the EOW branes in the bulk region. The planar replica wormholes with $l$-dotted loops and $m_b$-disconnected disk regions with $b$-asymptotic boundaries provide the factors $k^l \prod_{m_b, b} (Z_{b})^{m_b}$ respectively.}
\label{fig:EOW_model}
\end{figure}

In this section, we consider a planar approximation of the replica wormhole calculation in a simple 2$d$ gravity model consisting of a black hole in the JT gravity, a brane and an auxiliary system, discussed in \cite{Penington:2019kki} (dual to a pure state in the Sachdev-Ye-Kitaev model \cite{Kourkoulou:2017zaj}). 
To model the early radiation of an evaporating black hole, we consider a quantum state describing a black hole system $B$ of dimension $\text{dim}\, \CH_B = e^{S_0}$ with a brane entangled with an auxiliary system $R$ of dimension $\text{dim}\, \CH_R=k$\,:
\begin{align}
    | \Psi \rangle 
    = \frac{1}{\sqrt{k}} \sum_{i=1}^{k} | \psi_i \rangle_B\, | i \rangle_R\ . 
\end{align}
Here we take an orthogonal basis $| i \rangle_R$ of the auxiliary system $R$ and an unnormalized basis $| \psi_i \rangle_B$ of the black hole system. The $2d$ gravity system has two asymptotic $1d$ boundaries on $B$ and the brane (the left panel of figure \ref{fig:EOW_model}). The boundary conditions are given by the inverse temperature $\beta$ or the renormalized length on $B$ and the brane tension $\mu$.
As the black hole $B$ evaporates the auxiliary system $R$ will collect more radiated particles. 
In that sense, we can regard $\log (k\, e^{-S_0})$ as time in the model.

To discuss the island contributions in this model, 
we work on the replica method for the R\'enyi entropies $S_R^{(n)}$ in the auxiliary system $R$. 
The basic object we calculate is the $n^{th}$ moment of the reduced density matrix $\rho_R = \Tr_{B} |\Psi \rangle \langle \Psi |$ for $R$, 
\begin{align}
   \Tr\, \rho_R^n 
   = \frac{1}{(k\, e^{S_0})^n} \sum_{i_1, \cdots, i_n = 1}^{k}  
   \langle \psi_{i_1} | \psi_{i_2} \rangle_B \cdot 
   \langle \psi_{i_2} | \psi_{i_3} \rangle_B \cdots  \langle \psi_{i_{n}} | \psi_{i_1} \rangle_B\ .
\end{align}
This is divided by $e^{n S_0}$ so that $\Tr\, \rho_R = 1$. 
The products of the amplitudes $\prod_{k} \langle \psi_{i_k}| \psi_{i_{k+1}} \rangle_B$ is evaluated by the gravitational path integral on replicated manifolds with boundaries connecting some of the $n$-copies of the auxiliary system $R$.  
By taking the replica wormhole saddles in the gravitational path integral, 
and considering large $k, e^{S_0}$ with a fixed ratio $\alpha_0\equiv k\, e^{-S_0}$, we can pick up 
the replica wormholes with disk topology as dominant contributions (the right panel of figure \ref{fig:EOW_model}).
This is nothing but the planar approximation:
\begin{align} \label{eq:nth moment}
    \begin{aligned}
   \Tr\, \rho_R^n 
    &\approx 
    \frac{1}{(k\, Z_1)^n} \left[ k\, (Z_1)^n + \binom{n}{2} \cdot k^2 Z_2\, (Z_1)^{n-2} + \cdots + k^n Z_n \right]\\
    &= 
    \frac{1}{k^{n-1}} \left[ 1 + \binom{n}{2} \cdot \frac{k\, Z_2}{(Z_1)^2} + \cdots + \frac{k^{n-1} Z_n}{(Z_1)^n} \right]\ .
   \end{aligned}
\end{align}
where $Z_n$ is a partition function for a replica wormhole connecting $n$-copies of the auxiliary system $R$. 
Since the wormhole has disk topology, $Z_n \propto e^{S_0}$ and then the moment is a finite series of $\alpha_0$. 
The $n^{th}$ term $\propto \alpha_0^{n-1}$ corresponds to the fully connected replica wormhole with the coarse-grained or  black hole R\'enyi entropy $S_{\text{BH}}^{(n)} \equiv \frac{1}{1-n}\log \frac{Z_n}{(Z_1)^n}$ .

Although many types of the replica wormhole configurations contribute to the moment at intermediate order of $\alpha_0$, the entanglement entropy has two kinds of dominant phases in the asymptotic limits, the fully disconnected ($\alpha_0 \ll 1$) and fully connected replica wormhole phases ($\alpha_0 \gg 1$):
\begin{align}\label{eq:limSR}
    S_R 
    \approx  
    \begin{dcases}
    \log k & (\alpha_0 \ll 1) \ ,\\
    S_{\text{BH}} \equiv\lim_{n\to 1} S_{\text{BH}}^{(n)}
    & (\alpha_0 \gg 1)  \ .
  \end{dcases} 
\end{align}
Here we take the limit $n \to 1$ after picking up the leading contribution associated to $\alpha_0$. 
The entropy starts with a linear growth in $\log \alpha_0$ 
and ends up with the saturation to the large value of $ S_{\text{BH}}$ as the island formula suggests.
Note that, as we will see later in this section, 
the intermediate parts of the Page curves are smoothed out in this model.

Similarly, we can find the capacity of entanglement \eqref{def_capacity} in the asymptotic limits:
\begin{align}\label{cap_asympt}
    C_R 
    \approx  
    \begin{dcases} 
    k\,\frac{Z_2}{(Z_1)^2} \ \propto \alpha_0 & (\alpha_0 \ll 1) \ ,\\
    C_{\text{BH}}\equiv -2 \left.\partial_n S_{\text{BH}}^{(n)}\right|_{n=1} & (\alpha_0 \gg 1) \ .
  \end{dcases} 
\end{align}
At early time, the capacity grows exponentially in $\log \alpha_0$ from zero (or, if plotted in $\log k$, exponentially small initial value $e^{-S_0}$ for large $S_0$).
But, at late time, it decreases to the final value $C_{\text{BH}}$ which can be nonzero. 

Note that, although we cannot draw any implication about the behavior of the capacity around the Page time ($\alpha_0\sim 1$) from this simple discussion, 
as shown in the following subsections, the capacity takes the maximum around the Page time after the initial growth and then decreases exponentially to the late time value $C_{\text{BH}}$.
In semiclassical regime the capacity can be discontinuous due to the phase transition between the fully connected and fully disconnected replica wormholes, but the curve should be smoothed out once the other replica geometries are taken into account.
Even if the capacity does not exhibit a discontinuity it may still be a good indicator for the phase transition between the disconnected (no island) and connected replica wormhole (island) phases.

In the rest of this section, we will perform more detailed calculations of the entropy and capacity in the microcanonical and canonical ensembles.
They incorporate the contributions from planer replica wormholes which smooth out the sharp transition from the fully disconnected to the fully connected phases around the Page time. 

\subsection{Microcanonical ensemble}

To probe the entropy and capacity around the Page time, we need the explicit forms of the partition functions $Z_i~(i=1,\cdots, n)$ for the replica wormholes.
To this end, it will be convenient to use the density of states $D(\lambda)$ of the eigenvalue $\lambda$ for the reduced density matrix $\rho_R$,
\begin{align}
 D(\lambda) &= \lim_{\epsilon \to +0} \frac{1}{\pi}\, \text{Im} [ R (\lambda - \i\, \epsilon) ]\ ,
\end{align}
where $R(\lambda) =\sum_i R_{ii}(\lambda)$ is the trace of the resolvent matrix $R_{i j} (\lambda)$,  
\begin{align}
    R_{i j} (\lambda) 
    &= \left( \frac{1}{\lambda \mathbbm{1} - \rho_R } \right)_{i j}
    = \frac{1}{\lambda}\, \delta_{i j} + \sum_{n=1}^{\infty} \frac{1}{\lambda^{n+1}}\,(\rho_R^n)_{ij}\ , 
\end{align}
which can be determined via the Schwinger-Dyson equation for large $k$ and $S_0$ with fixed ratio $\alpha_0$ or in the planar limit. 
By using the density of states $D(\lambda)$, the entanglement entropy $S_R$ and capacity $C_R$ is expressed as integrals of the eigenvalues $\lambda$:
\begin{align}
    S_R &= -\int_0^\infty \d\lambda\, D(\lambda) \, \lambda\, \log \lambda \ ,  \label{eq:SRmc1} \\
    C_R &= \int_0^\infty \d\lambda\, D(\lambda) \, \lambda\, (\log \lambda)^2 - (S_R)^2 \ . \label{eq:CRmc1}
\end{align}

In the microcanonical ensemble, we fix the energy $E$ in the asymptotic region rather than the temperature or the renormalized length $\beta$.
In a small energy band around $E$ with the entropy $\mathbf{S} \sim \log$(number of states in the energy band), 
the Schwinger-Dyson equation for $R(\lambda)$ is simplified to a quadratic equation and the solution gives the density of states $D(\lambda)$ in the microcanonical ensemble \cite{Penington:2019kki}:
\begin{align} \label{eq:DoS}
    \begin{aligned}
 D(\lambda) 
 &= \frac{k\, e^{\mathbf{S}}}{2 \pi \lambda} \sqrt{\left(\lambda - \left(k^{-\frac{1}{2}} - e^{-\frac{\mathbf{S}}{2}} \right)^2 \right) \cdot \left( \left(k^{-\frac{1}{2}} + e^{-\frac{\mathbf{S}}{2}} \right)^2 - \lambda \right) } 
 + \delta (\lambda)\,  (k - e^{\mathbf{S}})\, \theta (k - e^{\mathbf{S}}) \\
 &= k^2 \cdot \left[ \frac{1}{2\pi \alpha \tilde{\lambda}} \sqrt{(\tilde{\lambda}-(1-\sqrt{\alpha})^2)\cdot ((1+\sqrt{\alpha})^2-\tilde{\lambda})} 
 + \delta(\tilde{\lambda}) \left(1- \frac{1}{\alpha}\right)\, \theta(\alpha-1) \right] \\
 &\equiv k^2 \cdot \tilde{D}(\tilde{\lambda})\ . 
 \end{aligned}
\end{align}
Here we rescaled the variables as $\tilde{\lambda} = k\, \lambda$ and denote the ratio of the dimensions of the systems in the energy band as $\alpha = k\, e^{-\mathbf{S}}$. 
For $(1-\sqrt{\alpha})^2 \leq \tilde{\lambda} \leq (1+\sqrt{\alpha})^2$, the first term in  $\tilde{D}(\tilde{\lambda})$ takes real values. And the second term localized at $\tilde{\lambda} = 0$ gives us the normalization of the density of states as follows, 
\begin{align}
    \begin{aligned}
    \int_0^{\infty} \d\lambda\, D(\lambda) 
        &= k \int_0^{\infty} \d\tilde{\lambda}\, \tilde{D}(\tilde{\lambda})= k\ , \\
    \int_0^{\infty} \d\lambda\, \lambda \, D(\lambda) 
        &= \int_0^{\infty} \d\tilde{\lambda}\, \tilde{\lambda}\, \tilde{D}(\tilde{\lambda}) = 1\ . 
    \end{aligned}
\end{align} 
For $\alpha = k\, e^{-\mathbf{S}} < 1$, we have $k$-states distributed over $(1-\sqrt{\alpha})^2 \leq \tilde{\lambda} \leq (1+\sqrt{\alpha})^2$. 
For $\alpha > 1$, we have $e^{\mathbf{S}}$-states in the same region and $(k-e^{\mathbf{S}})$-states at $\lambda=0$.  
The density of states is nothing but the entanglement spectrum of a subsystem of dimension $k$ in a random state of total dimension $k\, e^{\mathbf{S}}$ in the planar approximation \cite{Page:1993df}.
The capacity of entanglement for random pure states was also examined in \cite{deBoer:2018mzv}.

The density of states \eqref{eq:DoS} gives us express the entropy \eqref{eq:SRmc1} and the capacity \eqref{eq:CRmc1}, 
\begin{align}
    S_R 
    &= \mathbf{S}+\log \alpha + \int \d\tilde{\lambda}\, \tilde{D}(\tilde{\lambda})\,  (-\tilde{\lambda} \log \tilde{\lambda})\ , \\
    C_R  
     &= 
        \int \d\tilde{\lambda}\, \tilde{D}(\tilde{\lambda})\, \tilde{\lambda}\, (\log \tilde{\lambda})^2 - \left(\int \d\tilde{\lambda}\, \tilde{D}(\tilde{\lambda})\, \tilde{\lambda}\, \log \tilde{\lambda} \right)^2 \ ,
\end{align}
as integrals of $\tilde{\lambda}$.
As we will see soon, the entropy and capacity give us smooth curves in $\log \alpha = \log k -\mathbf{S}$ with the expected asymptotics.

In the case of the microcanonical ensemble, we can  compute analytically the R\'enyi entropy $S_{R}^{(n)}$, which is now given by
\begin{align}
    \begin{aligned} 
    S^{(n)}_{R} - \mathbf{S}
    &=  \log \alpha 
    + \frac{1}{1-n} \log \left[ \int_{(1-\sqrt{\alpha})^2}^{(1+\sqrt{\alpha})^2} \d\tilde{\lambda}\, \tilde{D}(\tilde{\lambda})\, \tilde{\lambda}^{n} \right]. \label{eq:Renyimc1}
    \end{aligned}
\end{align}
By changing the variable as $\lambda'= (\tilde{\lambda} - (1-\sqrt{\alpha})^2)/(4\sqrt{\alpha})$, which takes real values for $0 \leq \lambda' \leq 1$, we can express the $n^{th}$ moment by the hypergeometric functions:
\begin{align}\label{nth_moment}
    \begin{aligned}
     \int_{(1-\sqrt{\alpha})^2}^{(1+\sqrt{\alpha})^2} \d\tilde{\lambda}\, \tilde{D}(\tilde{\lambda})\, \tilde{\lambda}^{n}
    &= (1-\sqrt{\alpha})^{2(n-1)} \, \frac{8}{\pi} \, \int_{0}^{1} \d \lambda'\, \sqrt{\lambda' \, (1-\lambda')} \left( 1 + \frac{4\sqrt{\alpha}}{(1-\sqrt{\alpha})^2} \lambda' \right)^{n-1} \\
    &= (1-\sqrt{\alpha})^{2(n-1)} \, {}_{2}F_{1} \left(1-n,\,\frac{3}{2}\,;\,3;\, -\frac{4 \sqrt{\alpha}}{(1-\sqrt{\alpha})^2} \right) \\
    &= {}_{2}F_{1} \left(1-n,-n;2; \alpha \right) \\
    &= \alpha^{n-1} \, {}_{2}F_{1} \left(1-n,-n;2; \frac{1}{\alpha} \right) \ .
    \end{aligned}
\end{align}
Here we used some formulas for the linear and quadratic transformations of the hypergeometric functions.
Note that the $n^{th}$ moment \eqref{nth_moment} is the special case of the result for the random pure states obtained in \cite{deBoer:2018mzv}.
This is controlled by only one parameter $\alpha$ as expected in the planar limit.
For $0 \leq \alpha \leq 1$, the third expression in \eqref{nth_moment} is a convergent power series of $\alpha$ and it can be expanded in $(n-1)$ as follows:
\begin{align}
    \begin{aligned}
    &{}_{2}F_{1} \left(1-n,-n; 2; \alpha \right) \\
     &\quad 
     = 1+ (n-1)\, \frac{\alpha}{2} -  \frac{(n-1)^2}{2} \left[ 1 + \frac{3 \alpha}{2} + \frac{1-\alpha^2}{ \alpha} \log (1-\alpha) -2\,\text{Li}_2(\alpha)  \right] + O((n-1)^3)\ .
    \end{aligned}
\end{align}
For $1 \leq \alpha$, the last line of \eqref{nth_moment} gives us a similar expansion by replacing $\alpha$ to $1/\alpha$ in the above expansion. 
Then the R\'enyi entropy \eqref{eq:Renyimc1} is expanded as
\begin{align} \label{eq:Renyimc2}
    S^{(n)}_{R} -\mathbf{S}
    = \begin{dcases}
    \log \alpha -\frac{\alpha}{2} 
    - \frac{(n-1)}{2} \cdot \left[ -1 -\frac{3\alpha}{2} -\frac{\alpha^2}{4} -\frac{1-\alpha^2}{\alpha}\log(1-\alpha) +2\, \text{Li}_2 (\alpha) \right] +\cdots & (0 < \alpha \leq 1)\ , \\
     -\frac{1}{2\alpha} 
    - \frac{(n-1)}{2} \cdot \left[ -1 -\frac{3}{2\alpha} -\frac{1}{4\alpha^2} -\frac{\alpha^2-1}{\alpha}\log\left(\frac{\alpha-1}{\alpha} \right) +2\, \text{Li}_2 \left(\frac{1}{\alpha} \right) \right] + \cdots & (1 \leq \alpha)\ .
    \end{dcases}
\end{align}

\begin{figure}
    \centering
    \includegraphics[height=4.5cm]{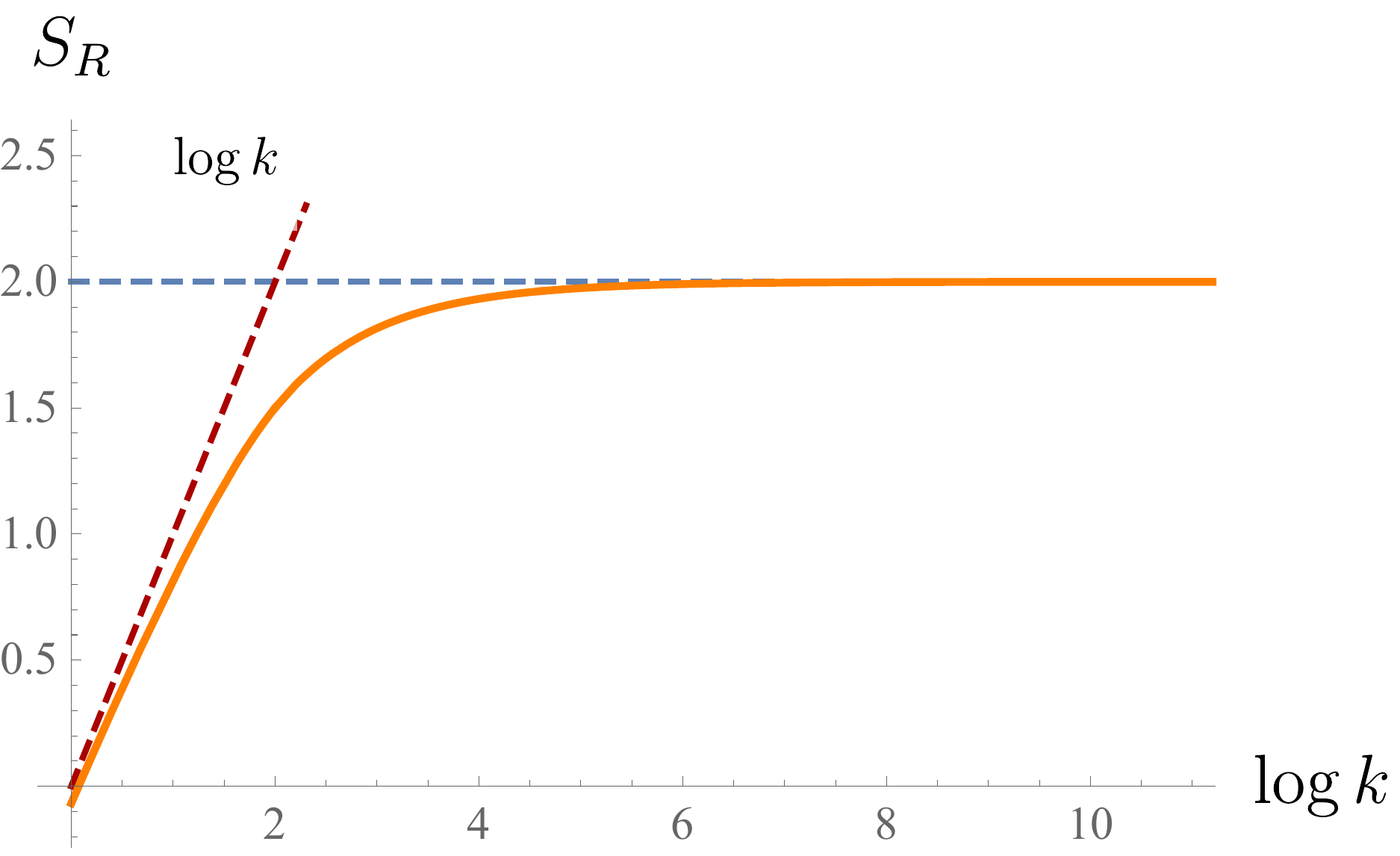}
    \qquad
    \includegraphics[height=4.5cm]{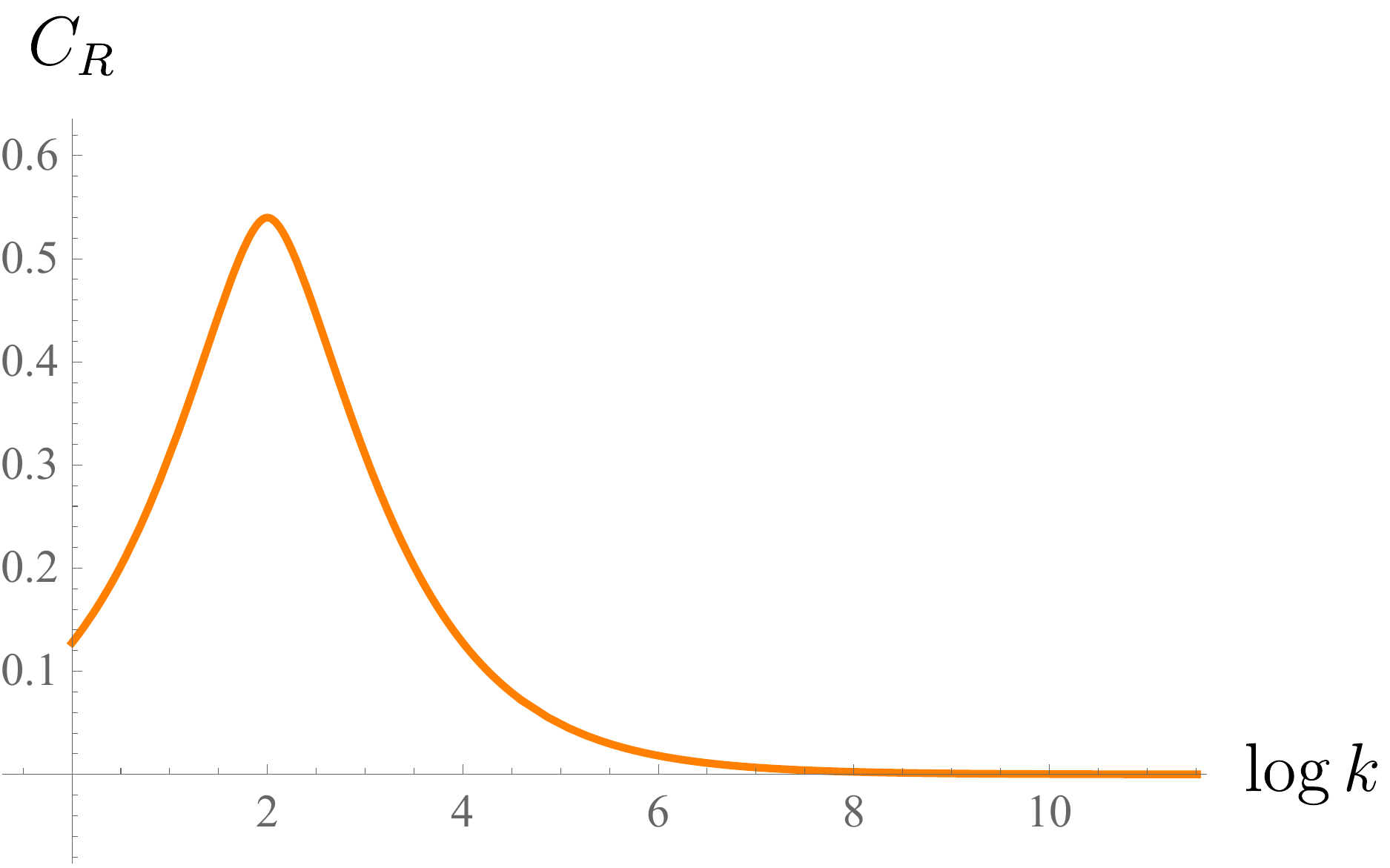}
    \caption{The entanglement entropy $S_R$ \eqref{eq:SRmc2} [Left] and capacity of entanglement $C_R$ \eqref{eq:CRmc2} [Right] for the EOW brane model with $\mathbf{S} = 2$ in the microcanonical ensemble.
    The capacity has a peak at the Page time $\log k = \mathbf{S}$, which clearly shows the crossover from the fully disconnected to fully connected replica wormhole solutions describing the Hawking radiation in an evaporating black hole.
    }
    \label{fig:QM_model}
\end{figure}

Taking the limit $n \to 1$ in \eqref{eq:Renyimc2}, we can find the entanglement entropy $S_R$,  
\begin{align} \label{eq:SRmc2}
    S_{R} -\mathbf{S}
    = \lim_{n \to 1} ( S_{R}^{(n)} -\mathbf{S} )
    =
    \begin{dcases}
    \log  \alpha -\frac{\alpha}{2} & (0 < \alpha \leq 1) \ ,\\
     -\frac{1}{2\alpha} & (1 \leq \alpha )\ ,
    \end{dcases}
\end{align}
which is smooth around the Page time ($\alpha \sim 1$) 
and has the desired asymptotic behaviors, $S_R \approx \{ \log k \,  (\alpha \ll 1), \mathbf{S} \, (\alpha \gg 1) \}$ (see the left plot in figure \ref{fig:QM_model}). 

Finally we can find a smooth curve of the capacity of entanglement $C_R$ from the first order coefficient in \eqref{eq:Renyimc2}:
\begin{align} \label{eq:CRmc2}
    C_{R} &= -2 \left. \partial_n S_R^{(n)} \right|_{n=1}
    = 
    \begin{dcases}
    -1 -\frac{3\alpha}{2} -\frac{\alpha^2}{4} -\frac{1-\alpha^2}{\alpha}\log(1-\alpha) +2\,\text{Li}_2(\alpha) & (0 < \alpha \leq 1)\ , \\
     -1 -\frac{3}{2\alpha} -\frac{1}{4\alpha^2} -\frac{\alpha^2-1}{\alpha}\log\left(\frac{\alpha-1}{\alpha} \right) +2\,\text{Li}_2\left(\frac{1}{\alpha} \right) & (1 \leq \alpha )\ .
    \end{dcases}
\end{align}
As shown in the right plot of figure \ref{fig:QM_model}, it has a peak around the Page time ($\alpha \sim 1$ or $\log k \sim \mathbf{S}$) signaling the phase transition. 
In the asymptotic regions, the capacity decays to zero polynomially in $\alpha$ or exponentially in $\log \alpha$:
\begin{align} \label{eq:CRmc3}
    C_{R} \sim 
    \begin{dcases}
    \frac{\alpha}{2} & (\alpha \ll 1)\ , \\
    \frac{1}{2\alpha} & (1 \ll \alpha )\ .
    \end{dcases}
\end{align}
The capacity becomes maximum at the Page time ($C_{R, \text{max}} = \frac{\pi^2}{3} -\frac{11}{4}\approx 0.54$) and decays to zero at the late time.
Note that, since the capacity $C_R$ is defined via replica parameter $n$ not temperature,
the late time value $C_\text{BH}$ introduced in \eqref{cap_asympt} is well-defined even in the microcanonical ensemble, and does not necessarily equal to thermal capacity which we will see in the canonical ensemble.

From these observations we conclude that the sharp peak at the Page time is a good diagnostics of the phase transition between the fully connected and fully disconnected replica wormholes.

\subsection{Canonical ensemble}

We consider the canonical ensemble by fixing the inverse temperature $\beta$ of the system rather than the energy. 
In the present model, $\beta$ corresponds to the renormalized length at the asymptotic boundary.
Since $\beta$ appears with $G_N$ as $G_N \beta$ and we consider small $G_N$ or the semi-classical regime, we will take small $\beta$.

In the canonical ensemble, the density of state may not be simplified analytically.
So we will try to analyze the resolvent  equation \eqref{eq:SDeq2} numerically with some approximation. (For the detail analysis, see \cite{Penington:2019kki}.) 

\begin{figure}
    \centering
    \includegraphics[height=4.5cm]{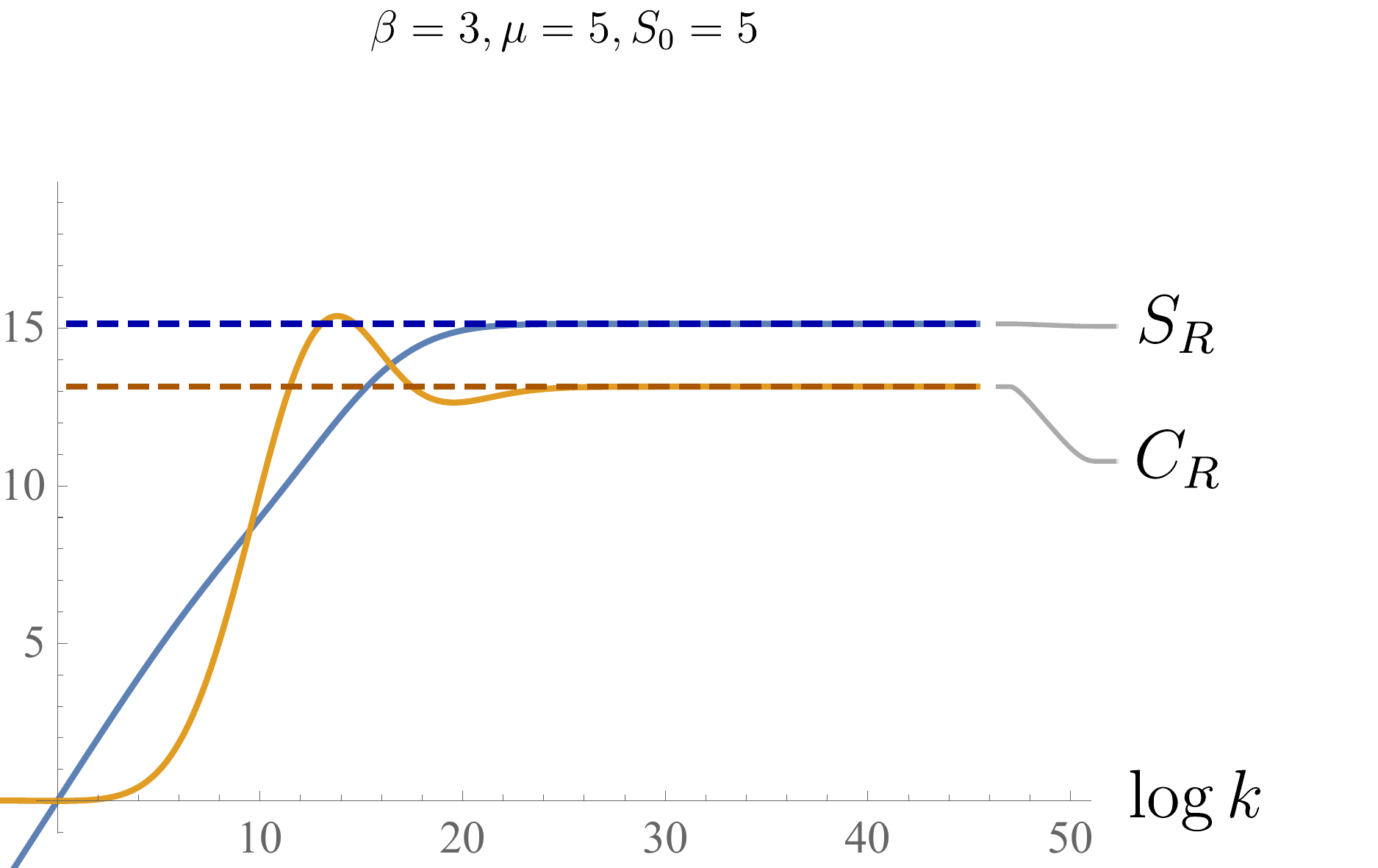}
    \quad
    \includegraphics[height=4.5cm]{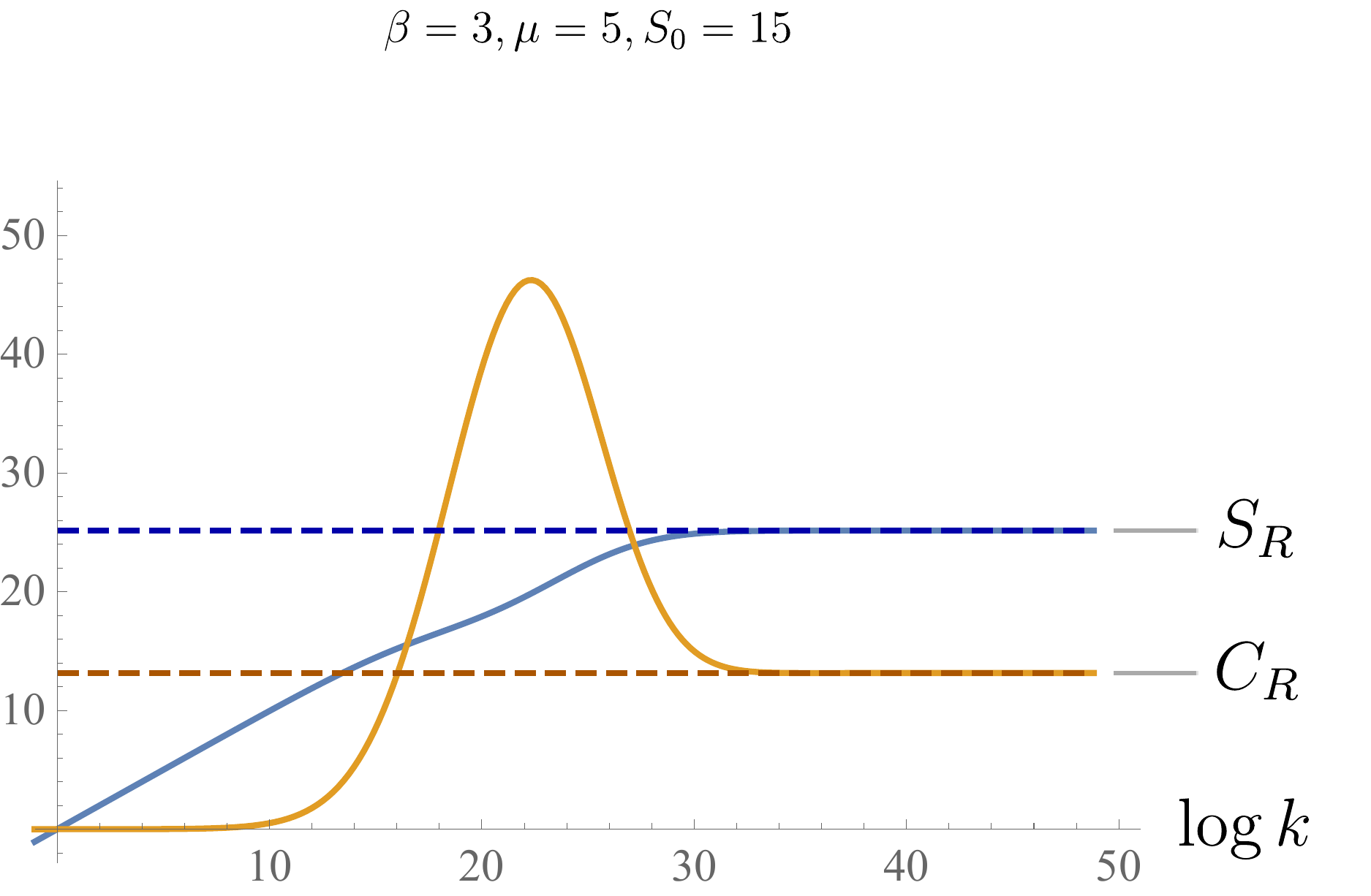}
    \caption{The entropy (blue) and capacity (orange) of the EOW brane model in the canonical ensembles.
    Both approach the asymptotic values $S_\text{BH}$ and $C_\text{BH}$ shown in the dashed lines in the large $k$ limit.
    }
    \label{fig:canonical}
\end{figure}

To calculate the resolvent $R(\lambda)$, we use the Schwinger-Dyson equation.  
Taking the planar limit (large $k$ and $S_0$ with $\alpha_0 = k\, e^{-S_0}$ fixed) and rewriting the replica partition functions $Z_n$ as integrals of the energy $s=\sqrt{2E}$, the equation ends up with a geometric series:
\begin{align} \label{eq:SDeq2}
    \lambda\, R (\lambda) = k + e^{S_0} \int_0^{\infty} \d s\, \rho(s)\, \frac{w(s) R (\lambda)}{k - w(s) R (\lambda)}\ , 
\end{align}
where 
\begin{align}
    \begin{aligned}
    \rho(s) &\equiv \frac{s}{2 \pi^2}\, \sinh (2 \pi s)\ , & \qquad
    y(s) &\equiv e^{-\frac{\beta s^2}{2}}\, 2^{1-2\mu}\, \left| \Gamma \left(\mu-\frac{1}{2} + \i\,s \right) \right|^2 \ , \\
    w(s) &\equiv \frac{y(s)}{Z_1} \ ,& \qquad Z_n &\equiv e^{S_0}\int_0^{\infty} \d s\, \rho(s)\, y(s)^n\ , 
    \end{aligned}
\end{align}
and $\mu$ is the brane tension (or mass of a particle behind the horizon) and $\beta$ is the renormalized length at the asymptotic boundary (or the temperature of the black hole) in the dual JT gravity. 
To gain the physical intuition for the spectrum, we divide the spectrum into three regimes: pre-Page, near-Page and post-Page phases. 
The density of state $D(\lambda)$ is well-localized near $\lambda=1/k$ at early time (the pre-Page phase).
But, as the subsystem evolves, the distribution decays into the thermal spectrum which has a long tail in high eigenvalues (the near-Page and post-Page phases). 
In order to truncate the thermal tail, we introduce an energy cutoff $s_k$ such that 
\begin{align} \label{eq:kdef}
    k\, e^{-S_0}= \int_{0}^{s_k} \d s\, \rho (s) = \frac{\cosh(2\pi s_k)}{8\pi^4} \cdot \left( 2 \pi s_k  - \tanh(2\pi s_k)\right)\ .
\end{align}
The spectrum has a minimal eigenvalue $\lambda_0$ which is determined by $\d\lambda/\d R = 0$ and the resolvent equation \eqref{eq:SDeq2}:
\begin{align}
\lambda_0 \approx \frac{e^{S_0}}{k} \int_{s_k}^{\infty} \d s\, \rho(s)\, w(s) 
    = \frac{1}{k} \left[ 1 - e^{S_0} \int_0^{s_k} \d s\, \rho(s)\, w(s) \right]\ . 
\end{align}
Under the condition $k \ll e^{S_0}\int_{0}^{s_k} \d s\, \rho(s)\, w(s)/(\lambda -\lambda_0 -w(s))$ or $\lambda > \lambda_0 + w(s_k -\delta)$ with a small control parameter $\delta$, 
the density of state is approximated as 
\begin{align}
    D(\lambda) 
    = e^{S_0} \int_{0}^{s_k} \d s\, \rho(s)\, \delta(\lambda - \lambda_0 -w(s))\ . 
\end{align}
Using this density of state, we find the entanglement entropy $S_R$ and the capacity $C_R$:
\begin{align}
    S_R &= -e^{S_0} \int_{0}^{s_k} \d s\, \rho(s)\, (\lambda_0 + w(s))\, \log (\lambda_0 + w(s))\ , \\
    C_R &= e^{S_0} \int_{0}^{s_k} \d s\, \rho(s)\, (\lambda_0 + w(s))\, (\log (\lambda_0 + w(s)))^2 - (S_R)^2\ . 
\end{align}
While the analytic forms of $S_R$ and $C_R$ are out of our reach, it is straightforward to perform these integrals numerically.
We plot them as functions of $\log k$ in figure \ref{fig:canonical}.

We can compare the numerical results with the analytic results in limits.
In the small $k$ region where $D(\lambda)$ is localized near $\lambda =1/k$, we find \cite{Penington:2019kki}
\begin{align}
    \begin{aligned}
        S_R 
        &= \log k + O(k\,e^{-S_0}) \ , \\
        C_R &=  O(k\,e^{-S_0}) \ .
    \end{aligned}
\end{align}
This explains the linear and exponential growths of $S_R$ and $C_R$ at the early time respectively. 

In the large $k$ region, by iterative uses of the resolvent equation \eqref{eq:SDeq2} for $\lambda \gg w(s_k)$, we can approximate the resolvent $R(\lambda)$ by the first order perturbation of $1/k$:
\begin{align}
    \begin{aligned}
    R(\lambda) 
        &\approx
        e^{S_0}\, \int_0^{s_k} \d s\, \rho(s)\, \frac{1}{\lambda-w(s)}\ .
    \end{aligned}
\end{align}
Here we also used the definition of $k$ \eqref{eq:kdef}. 
Then the density of state has the form of the thermal spectrum (consistent with $\lambda_0 \approx 0$ in large $k$):
\begin{align}
    D(\lambda) = e^{S_0}\, \int_0^{s_k} \d s \, \rho (s)\, \delta(\lambda - w(s))\ . 
\end{align}
Indeed it turns out that $D(\lambda) = 0$ for $\lambda \gg w(s_k)$, and there is no analytic control for the intermediate values.
The entanglement entropy approaches to the coarse-grained or black hole entropy $S_{\text{BH}}$ in the canonical ensemble:
\begin{align}
    S_{\text{BH}} 
    = -\frac{e^{S_0}}{Z_1}\int_{0}^{\infty} \d s\, \rho (s)\, y(s)\, \log y(s) + \log Z_1\ .  
\end{align}
For $\mu \gg 1/\beta$, $y(s) \sim y(0)\, e^{-\beta s^2 /2}$ since $|\Gamma(\mu -1/2 +\i\,s)|^2/(\Gamma(\mu-1/2))^2 = \prod_{m=0}^{\infty}(1+s^2/(m+\mu-1/2)^2)^{-1} \sim 1$.
Then the disk partition function and the entropy can be approximated as 
\begin{align}
    \begin{aligned}
        Z_1 &\approx \frac{e^{S_0 + 2\pi^2/\beta}}{\sqrt{2\pi}\, \beta^{3/2}} \ ,\\
        S_{\text{BH}} &\approx S_0 + \frac{4\pi^2}{\beta} + \frac{3}{2} \log \frac{2\pi}{\beta} + \frac{3}{2} -2\log(2\pi) \ .
    \end{aligned}
\end{align}
Similarly, the capacity of entanglement approaches to the black hole capacity $C_{\text{BH}}$ in the canonical ensemble:
\begin{align}
    \begin{aligned}
    C_{\text{BH}} 
    &= \frac{e^{S_0}}{Z_1} \int_{0}^{\infty} \d s\, \rho (s)\, y(s) \, (\log y(s))^2 -\left(\frac{e^{S_0}}{Z_1}\int_{0}^{\infty} \d s\, \rho (s) \, y(s)\, \log y(s) \right)^2\ .
\end{aligned}
\end{align}
For $\mu \gg 1/\beta$, the capacity approaches to 
\begin{align}
    C_{\text{BH}} \approx  \frac{4\pi^2}{\beta} + \frac{3}{2}\ ,
\end{align}
which is nonzero in contrast to the capacity \eqref{eq:CRmc2} in the microcanonical ensemble.

It might be worthwhile to mention that, although the capacity of entanglement $C_R$ is defined by using the derivatives of the replica parameter $n$ not the temperature $\beta$, interestingly the capacity in the canonical ensemble coincides with thermodynamic one $C_{\text{BH}}$ at late time.

\section{Moving mirror model of Hawking radiation}\label{ss:MovingMirror}
We consider another model of Hawking radiation known as a moving mirror model in two-dimensional CFT, where the mirror plays a role of a black hole horizon and a thermal energy flux can be measured at the infinity \cite{Davies:1976hi,birrell1984quantum,Carlitz:1986ng,Carlitz:1986nh,Wilczek:1993jn}.
The moving mirror may also be regarded as the EOW brane and 
this model has a similarity to the EOW brane model.
The entanglement entropy of a particular class of the moving mirror model has been investigated \cite{Bianchi:2014qua,Hotta:2015huj,Good:2016atu,Chen:2017lum,Good:2019tnf} and shown to reproduce the Page curves for eternal and evaporating black holes recently in \cite{Akal:2020twv}.
In this section we will extend the analysis of \cite{Akal:2020twv} to the capacity of entanglement.

\subsection{Moving mirror in CFT}

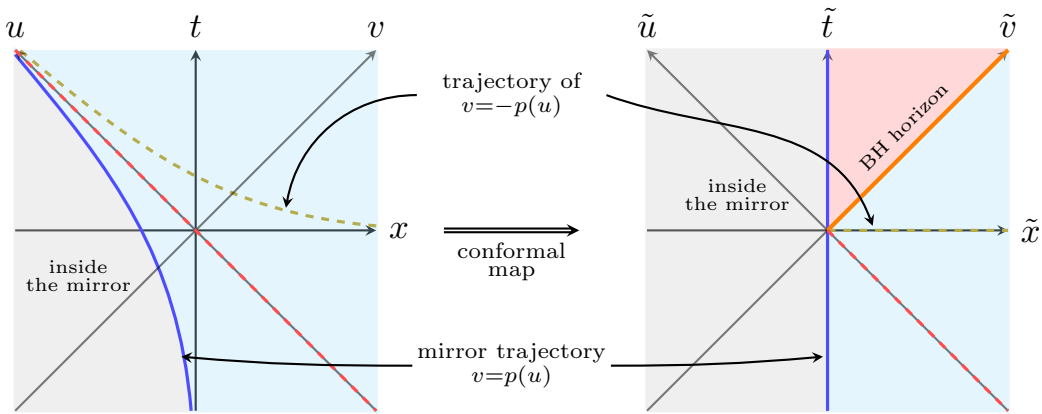
\begin{figure}
	\centering
	\begin{tikzpicture}[transform shape,scale=1.2]
	 \draw[thick,->,>=stealth] (-2,0) to (2,0) node [right, font=\normalsize] {$x$};
	            \draw[thick,->,>=stealth] (0,-2) to (0,2) node [above, font=\normalsize] {$t$};
	   \filldraw[very thick, gray!30,opacity=0.4] (-0.05,-2) to[out=95,in=-50] (-2,1.95) to (-2,-2) to (-0.05,-2);
	   \filldraw[very thick, cyan!30,opacity=0.3]  (2,2) to (2,-2) to (-2,2);
    \filldraw[very thick, cyan!30,opacity=0.3] (-0.05,-2) to[out=95,in=-50] (-2,1.95) to (-2,2) to (2,-2) to (-0.05,-2);
	   \draw[very thick, blue!70] (-0.05,-2) to[out=95,in=-50] (-2,1.95);
	   \draw[very thick, olive!70, dashed] (2,0.05) to[out=175,in=-40] (-1.95,2);

	   \draw[thick, black!100,->,>=stealth,opacity=0.5] (2,-2) to (-2,2) node [above, font=\normalsize,opacity=1] {$u$};
	   	   \draw[thick, black!100,->,->,>=stealth,opacity=0.5] (-2,-2) to (2,2) node [above, font=\normalsize,opacity=1] {$v$};
	   \draw[thick,double, black!100,->,>=stealth] (2.75,0) to (4.25,0);
	   \node[very thick,font=\scriptsize] at (-1.3,-0.5) {$\substack{\text{inside}\\\text{the mirror}}$};

	   \node[below, font=\small] at (3.5,0) {$\substack{\text{conformal}\\ \text{map}
	   }$};

	   \draw[very thick, red!70, dashed] (-2,2) to (2,-2);

\begin{scope}[shift={(7,0)}]
 \draw[thick,->,>=stealth] (-2,0) to (2,0) node [right, font=\normalsize] {$\tilde{x}$};
	            \draw[thick,->,>=stealth] (0,-2) to (0,2) node [above, font=\normalsize] {$\tilde{t}$};
	   \filldraw[very thick, gray!30,opacity=0.4] (-2,-2) to (0,-2) to (0,2) to (-2,2) to (-2,-2);
	   \filldraw[very thick, cyan!30,opacity=0.3] (2,2) to (2,-2) to (0,0);
	   \filldraw[very thick, red!30,opacity=0.5] (2,2) to (0,2) to (0,0);
    \filldraw[very thick, cyan!30,opacity=0.3] (2,-2) to (0,-2) to (0,0);
	   \draw[very thick, blue!70] (0,-2) to (0,2);
	   \draw[thick, black!100,->,>=stealth,opacity=0.5] (2,-2) to (-2,2) node [above, font=\normalsize,opacity=1] {$\tilde{u}$};
	   	   \draw[thick, black!100,->,->,>=stealth,opacity=0.5] (-2,-2) to (2,2) node [above, font=\normalsize,opacity=1] {$\tilde{v}$};
	   \node[very thick,font=\scriptsize] at (-1,0.4) {$\substack{\text{inside}\\\text{the mirror}}$};
	   \node[very thick,font=\tiny,above,rotate=45] at (1,1) {$\text{BH horizon}$};
	   \draw[very thick, olive!70, dashed] (0,0) to (2,0);
	   	   \draw[ultra thick, orange] (0,0) to (2,2);
	   	   	   	   \draw[very thick, red!70, dashed] (0,0) to (2,-2);
\end{scope}

	   \node[rectangle,
	   very thick,font=\small] at (3.5,1.5) {$\substack{\text{trajectory of}\\ v=-p(u)
	   }$};
	    \node[rectangle,
	    very thick,font=\small] at (3.5,-1.5) {$\substack{\text{mirror trajectory}\\ v=p(u)
	    }$};
	    \draw[thick, black!100,->,>=stealth] (2.4,-1.5) to[out=180,in=-5] (-0.16,-1.4);
	    \draw[thick, black!100,->,>=stealth] (4.6,-1.5) to[out=0,in=185] (6.95,-1.4);
	   \draw[black!100, thick,->,>=stealth] (2.45,1.5) to[out=180,in=80] (1,0.25);
	   \draw[black!100, thick,->,>=stealth] (4.55,1.5) to[out=-20,in=110] (7.5,0.05);

	\end{tikzpicture}
	\caption{A moving mirror model in the $(t,x)$-coordinate [Left] and in the tilde coordinates [Right].
	A CFT lives outside the mirror $v\geq p(u)$ ($x\geq z(t)$), which can be mapped to the RHP by a conformal transformation. The red region $(\tilde{x}\geq 0,\,\tilde{t}\geq \tilde{x})$ in the right panel has no counterpart in the original coordinate system and can be seen as a black hole interior.
	}
	\label{fig:moving mirror trajectory general}
\end{figure}

Suppose the trajectory of a mirror profile is given by $x=z(t)$.
We consider a CFT living in the region $x \geq z(t)$ and map it by a conformal transformation:
\begin{align}
\label{eq:mmconfmap}
 \tilde{u} = p(u)\ , \qquad \tilde{v}= v\ ,  
\end{align}
where $u=t-x$ and $v= t+x$.
In addition, we define a new coordinate after the conformal map $\tilde{x}$ and $\tilde{t}$ through the relation $\tilde{u}=\tilde{t}-\tilde{x}$ and $\tilde{v}=\tilde{t}+\tilde{x}$.
We choose the function $p(u)$ such that the mirror trajectory is mapped into a static one $\tilde{u}- \tilde{v} =0$ or equivalently $v = p(u)$, which can be written in the original coordinates as
\begin{align}
\label{eq:Zandp}
 t+ z(t) = p \left( t -z(t) \right)\ ,
\end{align}
as in figure \ref{fig:moving mirror trajectory general}.
More explicitly $p(u)$ can be written as
\begin{align}\label{mirror_traj}
    p(u) = 2\tau_u - u \ ,
\end{align}
where $\tau_u$ is subject to the condition $\tau_u - z(\tau_u) = u$.
Then, after the conformal map $p(u)$, CFT lives in the right half plane (RHP) $\tilde{x}\geq 0$.
If CFT is in the vacuum state in the tilde coordinates, the stress tensor in the original coordinates can be given by the Schwarzian term from the conformal anomaly:
\begin{align}\label{energy_flux}
    \langle\, T_{uu}\,\rangle = - \frac{c}{24\pi}\,\{ p(u), u\} \ ,
\end{align}
which can be matched to the energy flux of the Hawking radiation by an appropriate choice of the mirror trajectory $p(u)$ \cite{Davies:1976hi}.

Typically the mirror describing a (non-evaporating) black hole extends into the $x<0$ region, starting from $x=0$ at early time and asymptoting to the lightlike curve $v = 0$ at late time.
This means that the function $p(u)$ is subject to the following conditions:
\begin{align}\label{mirror_trajectory_condition}
    p(u)<0 \ , \qquad 
        p(u) = 
            \begin{dcases}
                u & \qquad (u\to -\infty)\ , \\
                -\beta e^{-u/\beta} & \qquad (u\to \infty) \ .
            \end{dcases}
\end{align}
Here $\beta$ corresponds to the inverse temperature of the black hole.
Also it follows from \eqref{mirror_traj} that 
\begin{align}\label{mirror_trajectory_condition2}
    p'(u) = \frac{1 + z'(t)}{1 - z'(t)} \ge 0 \ ,
\end{align}
unless the mirror travels faster than light.
We probe a CFT with this mirror trajectory by a semi-infinite line $R=[x_0,\infty)$ with $x_0>0$ as a radiation region.
In this case, the calculation of the entropy and capacity of entanglement amounts to evaluating the one-point function of the twist operator on the right half plane $\tilde{x}\geq 0$ \cite{Calabrese:2004eu}:
\begin{align}
    \begin{aligned}
    \braket{\,\sigma _n(t,x_0)\,}_{\rm RHP} &=
    e^{(1-n)\,S_{\rm bdy}}\left|\frac{2\tilde{x}_0}{\tilde{\epsilon}}\right|^{-\frac{c}{12}\left(n-\frac{1}{n}\right)} \\
    &= e^{(1-n)\,S_{\rm bdy}}\left|\frac{t+x_0 -p(t-x_0)}{\epsilon \sqrt{p'(t-x_0)}}\right|^{-\frac{c}{12}\left(n-\frac{1}{n}\right)}\ ,
    \end{aligned}
\end{align}
where the UV cutoff $\tilde{\epsilon}$ is introduced in the half space $\tilde{x}\geq0$ which is related by $\tilde{\epsilon} = \sqrt{p'(u)}\,\epsilon$ to the cutoff $\epsilon$ in the original half space $x\geq z(t)$.
$S_{\text{bdy}} = \log \, \langle 0 | B \rangle$ is the boundary entropy determined by the overlap between the vacuum $| 0 \rangle$ and a conformal boundary state $| B \rangle$ characterizing the vacuum degeneracy of the CFT with a boundary \cite{Affleck:1991tk}.
It can be interpreted as the entropy of a black hole the moving mirror describes.

Through the above one-point function, the R\'{e}nyi entropy for the subsystem $R$ at time $t$ is given by
\begin{align}
    \begin{aligned}
    S^{(n)}_{R} = \frac{c}{12}\,\left(1+\frac{1}{n}\right) \log \left[ \frac{t+x_0 -p(t-x_0)}{\epsilon \sqrt{p'(t-x_0)}} \right] + S_{\text{bdy}}\ .
    \end{aligned}
\end{align}
We then immediately obtain the entanglement entropy for $R$ at time $t$:
\begin{align}
    S_{R} = \frac{c}{6}\, \log \left[ \frac{t+x_0 -p(t-x_0)}{\epsilon \sqrt{p'(t-x_0)}} \right] + S_{\text{bdy}}\ ,
\end{align}
and the capacity of entanglement
\begin{align}
C_R = -2\, \partial_n S_R^{(n)} \big|_{n=1} 
= S_R - S_{\text{bdy}}\ .
\end{align}

For the mirror trajectory satisfying \eqref{mirror_trajectory_condition} it follows that the capacity is constant at early time and grows linearly in $t$ at late time:
\begin{align}
    C_R = 
        \begin{dcases}
            \frac{c}{6}\,\log \left( \frac{2x_0}{\epsilon}\right) & \qquad (t \to -\infty) \ ,\\
            \frac{c}{12\beta}\,t +  \frac{c}{6}\,\log \left( \frac{t}{\epsilon}\right) & \qquad (t \to +\infty)\ .
        \end{dcases}
\end{align}
The entropy takes the same form as $C_R$ with a shift by $S_\text{bdy}$, so it does not reproduce the Page curve.
Hence, there are no phase transitions to be detected by the capacity in this model.

\subsection{Moving mirror in holographic CFT}
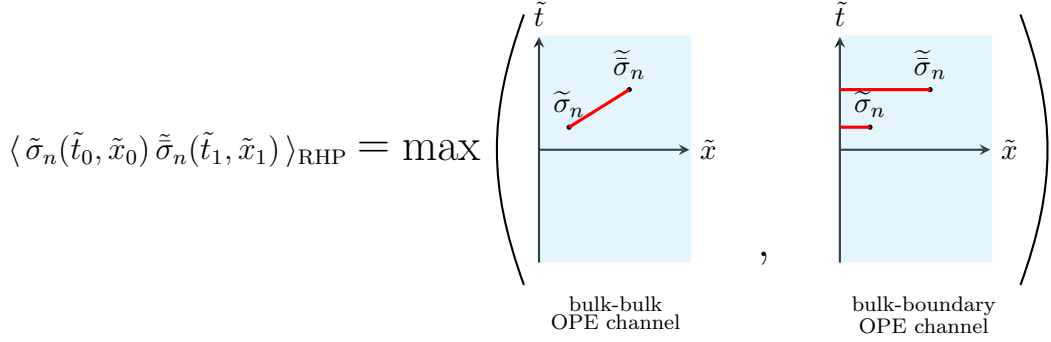
\begin{figure}
	\centering
	\begin{tikzpicture}[transform shape]
\begin{scope}[transform shape, scale=1,shift={(-2,0)}]
 	\draw[thick,->,>=stealth] (0,0) to (2,0) node [right, font=\normalsize] {$\tilde{x}$};
	\draw[thick,->,>=stealth] (0,-1.5) to (0,1.5) node [above, font=\normalsize] {$\tilde{t}$};
	\filldraw[very thick, cyan!30,opacity=0.3] (0,-1.5) to (0,1.5) to (2,1.5) to (2,-1.5);

    \filldraw[black!100, thick] (0.4,0.3) circle(0.02) node[above,black!100,thick] {$\widetilde{\sigma}_n$};

    \filldraw[black!100, thick] (1.2,0.8) circle(0.02) node[above,black!100,thick] {$\widetilde{\bar\sigma}_n$};

    \draw[red!100, very thick] (0.4,0.3) -- (1.2,0.8);
	   \node[black!100,ultra thick, font=\LARGE,below right] at (2.75,-1.2) {$,$};
	   
	 \node[black!100,ultra thick, font=\large,below right,rectangle] at (0,-1.8) {$\substack{\text{bulk-bulk}\\\text{OPE channel}}$};

    \node[black!100, very thick, font=\LARGE] at (-2.2,0) {$=$};
    \node[black!100, very thick, font=\LARGE] at (-1.3,0) {$\max$};	 
\end{scope}

\begin{scope}[transform shape, scale=1,shift={(2,0)}]
 	\draw[thick,->,>=stealth] (0,0) to (2,0) node [right, font=\normalsize] {$\tilde{x}$};
	            \draw[thick,->,>=stealth] (0,-1.5) to (0,1.5) node [above, font=\normalsize] {$\tilde{t}$};
	   \filldraw[very thick, cyan!30,opacity=0.3] (0,-1.5) to (0,1.5) to (2,1.5) to (2,-1.5);

    \filldraw[black!100, thick] (0.4,0.3) circle(0.02) node[above,black!100,thick] {$\widetilde{\sigma}_n$};

    \filldraw[black!100, thick] (1.2,0.8) circle(0.02) node[above,black!100,thick] {$\widetilde{\bar\sigma}_n$};

    \draw[red!100, very thick] (0.4,0.3) -- (0,0.3);
    \draw[red!100, very thick] (1.2,0.8) -- (0,0.8);

	 \node[black!100,ultra thick, font=\large,below right,rectangle
	 ] at (0,-1.8) {$\substack{\text{bulk-boundary}\\\text{OPE channel}}$};
	 
	 	\draw[thick] (-4.2,-1.8) to[out=110,in=-110] (-4.2,1.8);
	 	
		\draw[thick] (2.4,-1.8) to[out=70,in=-70] (2.4,1.8);
		
    \node[black!100, very thick, font=\large] at (-8.8,0) {$\langle\, \tilde{\sigma}_n (\tilde{t}_{0}, \tilde{x}_{0})\, \tilde{\bar{\sigma}}_n (\tilde{t}_{1}, \tilde{x}_{1})\, \rangle_{\text{RHP}}$};	 
\end{scope}
	\end{tikzpicture}
	\caption{The bulk-bulk OPE channel and the bulk-boundary OPE channel contributing to the two-point function of twist operators in two-dimensional holographic BCFT.
	A channel with larger value is favored, and there can be a phase transition between the two.
	}
	\label{fig:BCFT twist two point func at large c}
\end{figure}

Next we consider a moving mirror in a class of two-dimensional CFTs known as holographic CFTs with a large central charge.
We take the radiation region to be a finite subsystem $R = [x_0, x_1 ]$ at time $t$.
To obtain the R\'enyi entropy, we need to compute the two-point function of twist operators:
\begin{align}
\langle\, \sigma_n (t_{0}, x_{0})\, \bar{\sigma}_n (t_1, x_1)\, \rangle
= \left(p' (u_0)\, p' (u_1) \right)^{h_n} \cdot \langle\, \tilde{\sigma}_n (\tilde{t}_{0}, \tilde{x}_{0})\, \tilde{\bar{\sigma}}_n (\tilde{t}_{1}, \tilde{x}_{1}) \,\rangle_{\text{RHP}}\ .
\end{align}
In this case, the correlation function on the RHP can be expressed by two kinds of the identity blocks in different operator product expansion (OPE) channels of the two-point functions on $\mathbb{R}^{1, 1}$ in the large central charge limit $c \to \infty$ \cite{Sully:2020pza} (see figure \ref{fig:BCFT twist two point func at large c}):
\begin{align}
    \langle\, \tilde{\sigma}_n (\tilde{t}_{0}, \tilde{x}_{0})\, \tilde{\bar{\sigma}}_n (\tilde{t}_{1}, \tilde{x}_{1})\, \rangle_{\text{RHP}} = \max 
    \begin{dcases}
    \langle\, \tilde{\sigma}_n (\tilde{t}_{0}, \tilde{x}_{0})\, \tilde{\bar{\sigma}}_n (\tilde{t}_{1}, \tilde{x}_{1})\, \rangle_{\mathbb{R}^{1, 1}} \ ,\\
    e^{2(1-n) S_{\text{bdy}}} \cdot \prod_{i \in \{0, 1 \} }
    \langle\, \tilde{\sigma}_n (\tilde{t}_{i}, \tilde{x}_{i}) \, \tilde{\bar{\sigma}}_n (\tilde{t}_{i}, -\tilde{x}_{i})\, \rangle_{\mathbb{R}^{1, 1}}^\frac{1}{2}\ ,
    \end{dcases}
\end{align}
where the correlator of the twist operators on flat space is given by
\begin{align}
 \langle\, \tilde{\sigma}_n (\tilde{t}, \tilde{x}) \, \tilde{\bar{\sigma}}_n (\tilde{t}', \tilde{x}') \,\rangle_{\mathbb{R}^{1, 1}} 
 = \left| (\tilde{t} - \tilde{t}')^2 - (\tilde{x} - \tilde{x}')^2 \right|^{-\frac{c}{12} \left(n-\frac{1}{n}\right)}\ .
\end{align}
Whether a phase transition between the two channels occurs depends on both $S_\text{bdy}$ and the positions of the twist operators.

It is then straightforward to calculate the R\'enyi entropy:
\begin{align}
    \begin{aligned}
    S_R^{(n)}  
        &= \frac{1}{1-n} \log \, \langle \, \sigma_n (t, x_{0})\, \bar{\sigma}_n (t, x_1)\,  \rangle \\
        &= \min 
        \begin{dcases}
         \frac{c}{12}\,\left(1 + \frac{1}{n}\right)\, \log \left[ \frac{(x_1 - x_0) \left(p(t-x_0) - p(t-x_1)\right)}{\epsilon^2 \sqrt{p'(t-x_0)\, p'(t-x_1)}} \right] \ ,\\
         \frac{c}{12}\,\left(1 + \frac{1}{n}\right)\,\log \left[ \frac{t+x_0 -p(t-x_0)}{\epsilon \sqrt{p'(t-x_0)}} \right] + \frac{c}{12}\,\left(1 + \frac{1}{n}\right)\, \log \left[ \frac{t+x_1 -p(t-x_1)}{\epsilon \sqrt{p'(t-x_1)}} \right]  + 2 S_{\text{bdy}}\ .
         \end{dcases}
    \end{aligned}
\end{align}
In the $n \to 1$ limit, we can obtain the same entanglement entropy as the holographic one \cite{Takayanagi:2011zk,Fujita:2011fp}:
\begin{align}
    S_R =  S_R^{(n)} \big|_{n=1}= \min \left[S_{R}^{\text{dis}}, S_{R}^{\text{con}}\right]\ , 
\end{align}
where
\begin{align}
    \begin{aligned}
    \label{eq:mmentropy}
    S_{R}^{\text{dis}} 
        &= \frac{c}{6} \log \left[ \frac{t+x_0 -p(t-x_0)}{\epsilon \sqrt{p'(t-x_0)}} \right] + \frac{c}{6} \log \left[ \frac{t+x_1 -p(t-x_1)}{\epsilon \sqrt{p'(t-x_1)}} \right] + 2 S_{\text{bdy}}\ , \\
    S_{R}^{\text{con}} 
        &= \frac{c}{6} \log \left[ \frac{(x_1 - x_0) \left(p(t-x_0) - p(t-x_1)\right)}{\epsilon^2 \sqrt{p'(t-x_0)\, p'(t-x_1)}} \right]\ .
    \end{aligned}
\end{align}
When the conditions \eqref{mirror_trajectory_condition} and \eqref{mirror_trajectory_condition2} are met, 
the connected channel entropy $S_{R}^{\text{con}}$ is always favored at late time as the disconnected channel entropy grows as $S_{R}^{\text{dis}}\to \frac{c}{3}\,\log \left( t/\epsilon\right)$ in $t\to \infty$ while $S_{R}^{\text{con}}$ is bounded from above.
On the other hand, the dominant channel at early time depends on the value of the boundary entropy $S_\text{bdy}$.
When the disconnected channel is dominant at early time, $S_{R}^{\text{dis}} < S_{R}^{\text{con}}$, there must be a transition at the Page time where $S_R^{\text{con}} = S_R^{\text{dis}}$ and the entropy $S_R$ continuously changes its behavior from the linear growth in $\log t$ to plateau, reproducing the Page curve \cite{Akal:2020twv}.

The capacity of entanglement also depends on the dominant channel as given by
\begin{align}
\label{eq:mmcapacity}
    C_R = -2\, \partial_n S_R^{(n)} \big|_{n=1} = 
    \begin{dcases}
    S_R^{\text{con}} &\qquad \text{(connected channel)}\ ,\\
    S_R^{\text{dis}} -2S_{\text{bdy}} &\qquad \text{(disconnected channel)}\ .
    \end{dcases}
\end{align}
When the phase transition occurs for the entropy the capacity shows a discontinuity by $2 S_{\text{bdy}}$ at the Page time. 
This is a universal behavior independent of the profile of the moving mirror as long as the conditions \eqref{mirror_trajectory_condition} are met.

\begin{figure}[ht!]
	\centering
	\begin{tikzpicture}[transform shape,scale=1.2]
	 \draw[thick,->,>=stealth] (-2,0) to (2,0) node [right, font=\normalsize] {$x$};
	            \draw[thick,->,>=stealth] (0,-2) to (0,2) node [above, font=\normalsize] {$t$};
	   \filldraw[very thick, gray!30,opacity=0.6] (-0,-2) to[out=95,in=-60] (-0.3,0) to[out=120,in=-46] (-2,1.95) to (-2,-2) to (-0,-2);
	   \filldraw[very thick, cyan!30,opacity=0.3]  (2,2) to (2,-2) to (-2,2);
        \filldraw[very thick, cyan!30,opacity=0.3] (-0,-2) to[out=95,in=-60] (-0.3,0) to[out=120,in=-46] (-2,1.95) to (-2,2) to (2,-2) to (-0,-2);
	   \draw[very thick, blue!70] (-0.01,-2) to[out=91,in=-60] (-0.3,0) to[out=120,in=-46] (-2,1.95);

	   \draw[thick, black!100,->,>=stealth,opacity=0.5] (2,-2) to (-2,2) node [above, font=\normalsize,opacity=1] {$u$};
	   	   \draw[thick, black!100,->,->,>=stealth,opacity=0.5] (-2,-2) to (2,2) node [above, font=\normalsize,opacity=1] {$v$};

        \draw[very thick, |-|, magenta!70] (-0.8,1) -- (0.6,1) node[midway,above,very thick, black!100] {$R$~~~};
	\end{tikzpicture}
	\caption{A mirror trajectory describing an eternal black hole \eqref{eq:eternaltraj}. 
	It starts from $x=0$ and asymptotes to the light-ray $t= - x$ at late time.
	The entropy and capacity of entanglement are probed by the interval $R=[z(t)+0.1,z(t)+10]$ at a fixed distance from the mirror.}
	\label{fig:eternal moving mirror trajectory}
\end{figure}
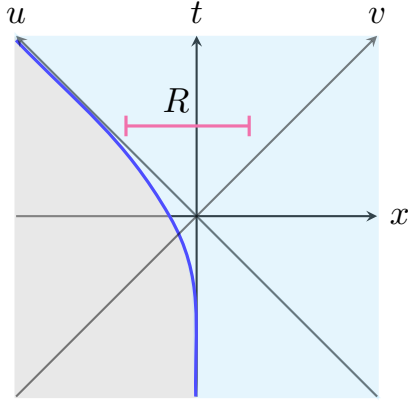

\paragraph{Eternal black hole}
As a concrete example, we consider the following moving mirror trajectory satisfying the conditions \eqref{mirror_trajectory_condition} and \eqref{mirror_trajectory_condition2}:
\begin{align}
\label{eq:eternaltraj}
  p(u) = - \beta\, \log \left(1 + e^{-u/\beta}\right) \ .
\end{align}
The stress tensor \eqref{energy_flux} is given by 
\begin{align}
    \langle\, T_{uu}\,\rangle = \frac{c}{48\pi\beta^2}\,\frac{1+ 2\,e^{-u/\beta}}{(1+e^{-u/\beta})^2} \ ,
\end{align}
which is zero at early time ($u\to -\infty$), but approaches to the energy flux of the Hawking radiation from a black hole at inverse temperature $\beta$ at late time ($u\to \infty$).
Thus this model is seen to describe an eternal black hole with radiation.
The original mirror trajectory $z(t)$ can be read off from $p(u)$ through the relation \eqref{eq:Zandp}.
For small $\beta$, the mirror trajectory $z(t)$ corresponding to the equation \eqref{eq:eternaltraj} can be approximated as follows:
at the early period $t < 0$, the mirror stays around the origin $z(t)= 0$ and approaches to the light-ray $z(t) = -t$ at the late time $t > 0$ as in figure \ref{fig:eternal moving mirror trajectory}.

\begin{figure}[ht]
    \centering
    \includegraphics[height=4.5cm]{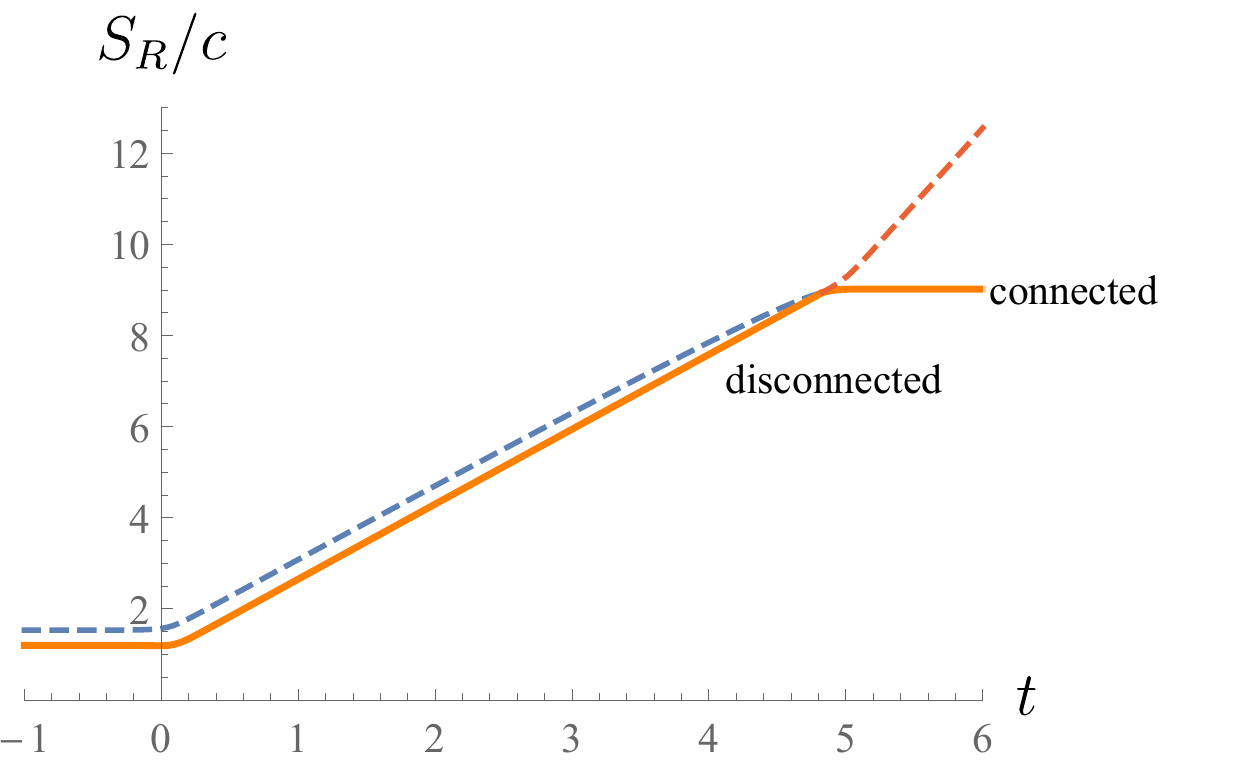}
    \quad
    \includegraphics[height=4.5cm]{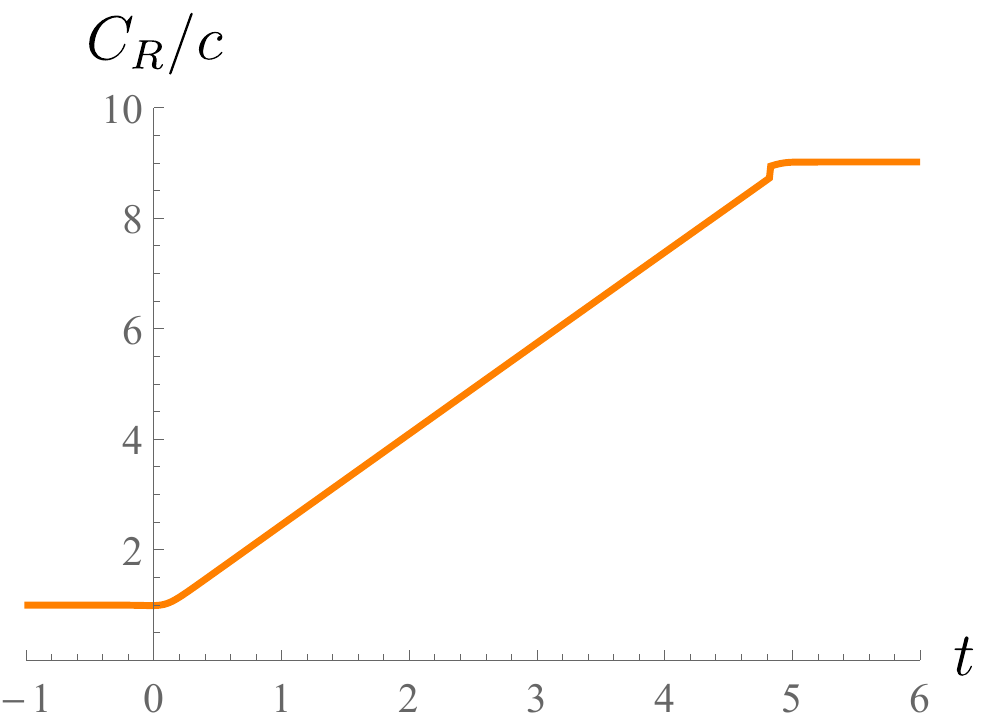}
    \caption{The entropy [Left] and capacity [Right] in the moving mirror model whose trajectory mimics an eternal black hole with radiation. We set the subsystem to be an interval $R = [z(t)+0.1, z(t)+10]$ with $\beta = 0.1$, $\epsilon = 0.1$ and $S_{\rm bdy}/c = 0.1$. The phase transition from the disconnected channel to connected one occurs around $t_\text{phase}=5$. The capacity shows a discontinuity by $2S_{\rm bdy} = 0.2c$ at that time.
    }
    \label{fig:MMeternal}
\end{figure}

We probe the entropy and capacity by the interval $R=[z(t)+0.1,z(t)+10]$ at time $t$.
The interval moves along the mirror trajectory with the distance kept fixed.
Using the equation \eqref{eq:mmentropy} with the boundary entropy $S_\text{bdy}/c = 0.1$, we plot the entanglement entropy in the left panel of figure \ref{fig:MMeternal}.
The disconnected channel is dominant at early time while the connected channel becomes dominant at late time.
The capacity of entanglement is obtained through the equation \eqref{eq:mmcapacity} as depicted in the right panel of figure \ref{fig:MMeternal}.
The discontinuity of the capacity at the phase transition time $t_{\rm phase}$,
\begin{align}
    \begin{aligned}
    C^{\rm con}-C^{\rm dis}\big|_{t_{\rm phase}} = 2S_{\rm bdy}\ ,
    \end{aligned}
\end{align}
is seen in the figure.

\begin{figure}
	\centering
	\begin{tikzpicture}[transform shape,scale=1.2]
	 \draw[thick,->,>=stealth] (-2,0) to (2,0) node [right, font=\normalsize] {$x$};
	            \draw[thick,->,>=stealth] (0,-2) to (0,2) node [above, font=\normalsize] {$t$};
	   \filldraw[very thick, gray!30,opacity=0.6] (-0,-2) to[out=90,in=-75] (-0.2,0) to[out=100,in=-50] (-0.85,0.85) to[out=110,in=-90] (-1,2) to (-2,2) to (-2,-2) to (-0,-2);
	   \filldraw[very thick, cyan!30,opacity=0.3]  (-0,-2) to[out=90,in=-75] (-0.2,0) to[out=100,in=-50] (-0.85,0.85) to[out=110,in=-90] (-1,2) to (2,2) to (2,-2) to (-0,-2);
	   \draw[very thick, blue!70] (-0,-2) to[out=90,in=-75] (-0.2,0) to[out=100,in=-50] (-0.85,0.85) to[out=110,in=-90] (-1,2) ;
	   \draw[thick, red!100, dotted, opacity=0.6] (-1,2) -- (-1,-2) node[below, black, ultra thick,font=\small, opacity=1] {$x=-\frac{u_0}{2}$};

   \draw[thick, black!100,->,>=stealth,opacity=0.5] (2,-2) to (-2,2) node [above, font=\normalsize,opacity=1] {$u$};
	   	   \draw[thick, black!100,->,->,>=stealth,opacity=0.5] (-2,-2) to (2,2) node [above, font=\normalsize,opacity=1] {$v$};

    \draw[very thick, magenta!70, |-|] (-0.4,0.6) -- (0.9,0.6) node[midway,above,very thick, black!100] {$R$};
	\end{tikzpicture}
	\caption{The mirror trajectory describing an evaporating black hole \eqref{eq:evaporatingtraj}. 
	It starts from $x=0$, approaches to the light-ray $t=-x$ around the time interval $t = [0, u_0/2]$, then asymptotes to $x= - u_0/2$ at late time.
	The region probing the entropy and capacity of entanglement is located at $R=[z(t)+0.1,z(t)+10]$.}
	\label{fig:evaporating moving mirror trajectory}
\end{figure}
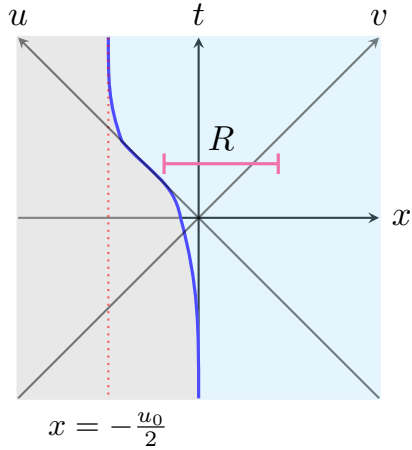

\paragraph{Evaporating black hole}
Next we consider another moving mirror trajectory \cite{Akal:2020twv}
\begin{align}\label{eq:evaporatingtraj}
  p(u) = - \beta\, \log \left(1 + e^{-u/\beta}\right)+ \beta\, \log \left(1 + e^{(u-u_0)/\beta}\right) \ .
\end{align}
In terms of the original coordinates $z(t)$ the mirror starts from the origin at early time, linearly decreases around the period $0 < t < u_0/2$ for $u_0>0$, then asymptotes to a constant $z(t) = -u_0/2$ at late time as in figure \ref{fig:evaporating moving mirror trajectory}.
Note that this trajectory does not satisfy the conditions \eqref{mirror_trajectory_condition} and \eqref{mirror_trajectory_condition2}, so the model describes not an eternal black hole but an evaporating one.
Indeed the energy flux 
\begin{align}
    \langle\, T_{uu}\,\rangle = \frac{c}{48\pi\beta^2}\left[\frac{1+ 2\,e^{-u/\beta}}{(1+e^{-u/\beta})^2} + \frac{1+ 2\,e^{(u-u_0)/\beta}}{(1+e^{(u-u_0)/\beta})^2}\right]
\end{align}
localized around finite time interval $0<t<u_0/2$ \cite{Akal:2020twv}, so the black hole is seen to evaporate at late time.
In contrary to the eternal case, the connected channel in this model does not necessarily dominate at late time.

\begin{figure}
    \centering
    \includegraphics[height=4.5cm]{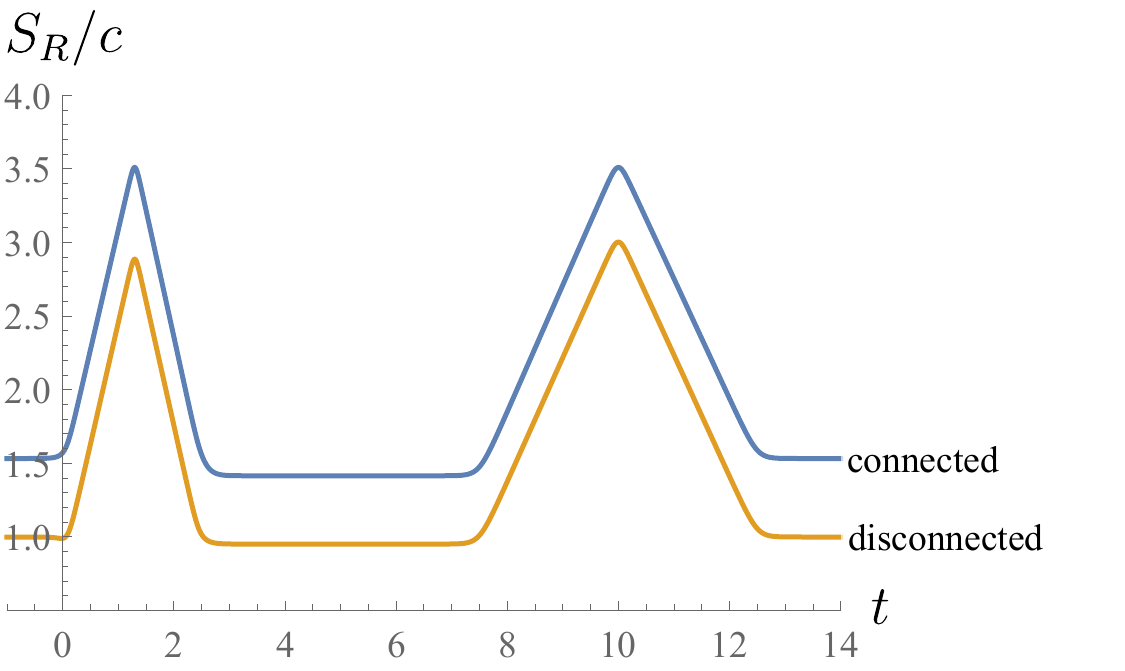}
    \quad
    \includegraphics[height=4.5cm]{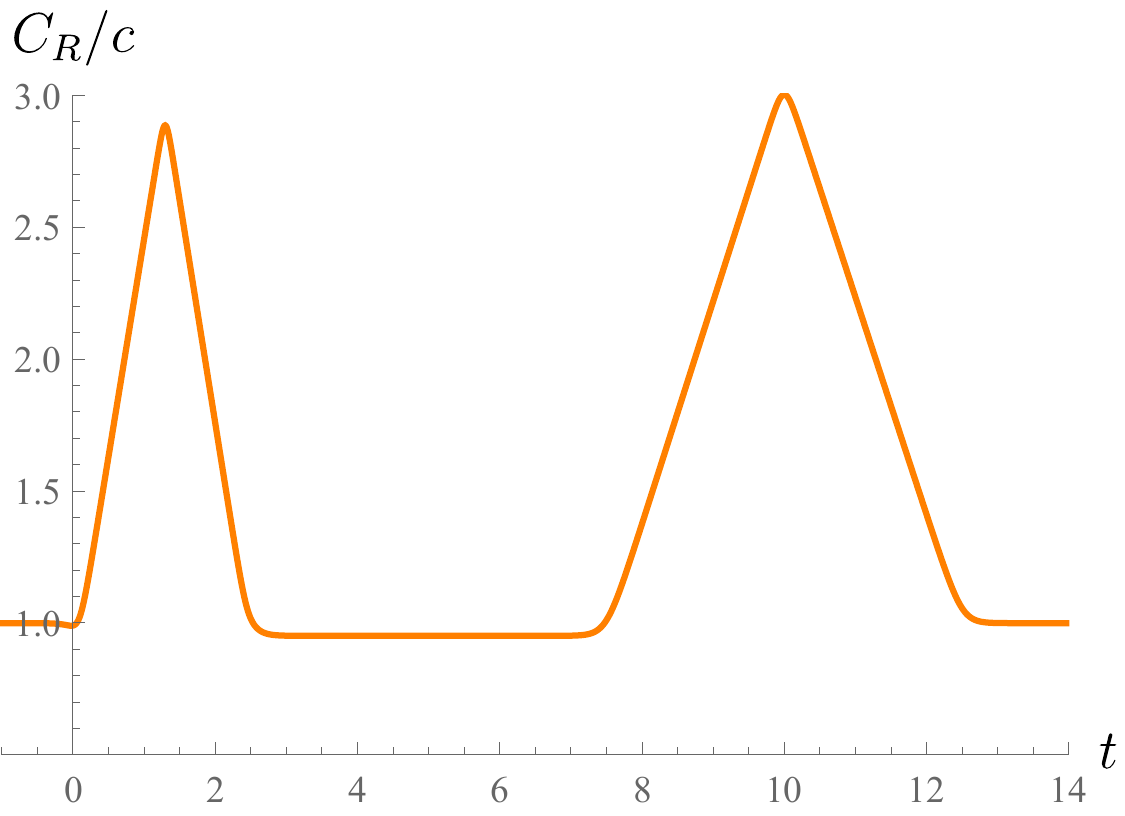}
    \caption{The entropy [Left] and capacity [Right] in the moving mirror model whose trajectory mimics an evaporating black hole. We set the subsystem to be $R = [z(t)+0.1, z(t)+10]$ with $\beta = 0.1$, $\epsilon = 0.1$ and $S_{\rm bdy}/c = 0.1$. In this model, there is no phase transition between the connected and disconnected channels, hence the capacity does not have any discontinuity.
    }
    \label{fig:MMevaporate}
\end{figure}

Similarly to the eternal case we probe the entropy and capacity by a finite interval $R = [z(t)+0.1, z(t)+10]$.
In this case the entropy reproduces the Page curve of an evaporating black hole, but with two peaks as in the left panel of figure \ref{fig:MMevaporate}.
While there are both connected and disconnected solutions, the disconnected one is always favored and no phase transition occurs between the channels in this model.
Correspondingly the capacity of entanglement does not show any discontinuity as seen in the right panel of figure \ref{fig:MMevaporate}.

\paragraph{Static mirror probed by an expanding interval}
Finally we consider a different setup from the previous ones, which has a static mirror but a radiation region expands in time.
While it does not describe a radiating black hole, it turns out that this model has a phase transition and may serve as a good testing ground for the capacity as an order parameter.

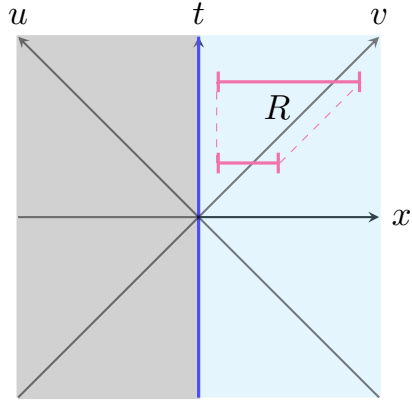
\begin{figure}
	\centering
	\begin{tikzpicture}[transform shape,scale=1.2]
	\draw[thick,->,>=stealth] (-2,0) to (2,0) node [right, font=\normalsize] {$x$};
	            \draw[thick,->,>=stealth] (0,-2) to (0,2) node [above, font=\normalsize] {$t$};
	   \filldraw[very thick, gray!60,opacity=0.6] (-2,-2) to (0,-2) to (0,2) to (-2,2) to (-2,-2);
	   \filldraw[very thick, cyan!30,opacity=0.3] (0,-2) to (0,2) to (2,2) to (2,-2);

	   \draw[very thick, blue!70] (0,-2) to (0,2);

	   \draw[thick, black!100,->,>=stealth,opacity=0.5] (2,-2) to (-2,2) node [above, font=\normalsize,opacity=1] {$u$};
	   	   \draw[thick, black!100,->,->,>=stealth,opacity=0.5] (-2,-2) to (2,2) node [above, font=\normalsize,opacity=1] {$v$};

        \draw[very thick, magenta!70, |-|] (0.2,0.6) -- (0.9,0.6);
        \draw[very thick, magenta!70, |-|] (0.2,1.5) -- (1.8,1.5) node[midway,below,very thick, black!100] {$R$~~~};
        \draw[dashed, magenta!70] (0.2,0.6) -- (0.2,1.5);
        \draw[dashed, magenta!70] (0.9,0.6) -- (1.8,1.5);
	\end{tikzpicture}
	\caption{The static moving mirror trajectory \eqref{eq:statictraj} probed by an expanding interval $R=[0.1, 0.5 + t]$.}
	\label{fig:static mirror traj general CFT}
\end{figure}

\begin{figure}
    \centering
    \includegraphics[height=4.5cm]{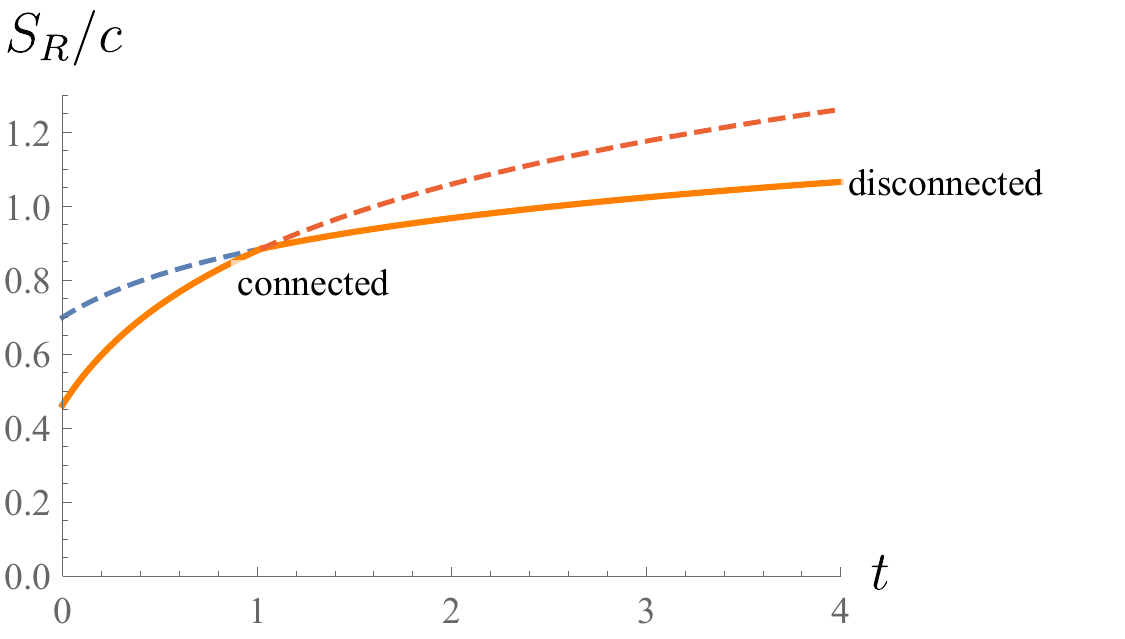}
    \quad
    \includegraphics[height=4.5cm]{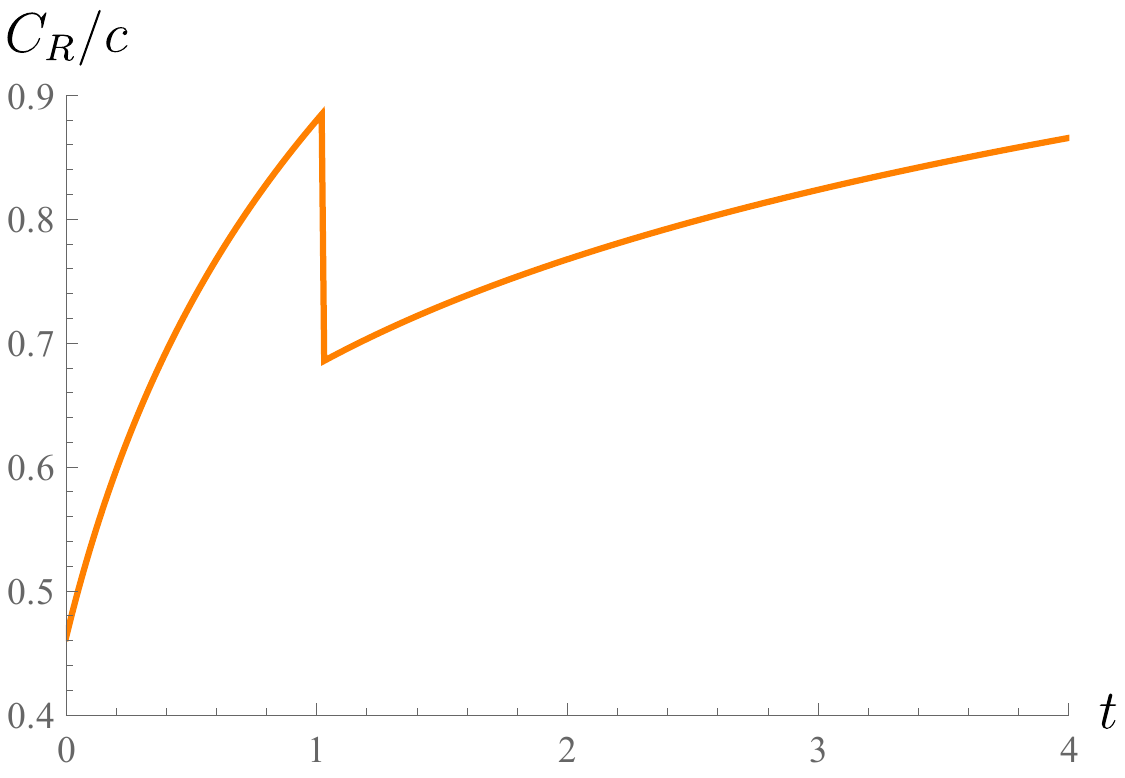}
    \caption{The entropy [Left] and capacity [Right] in the static mirror model probed by an expanding interval.  We set the subsystem to be $R = [0.1, 0.5+t]$ with $\epsilon = 0.1$ and $S_{\rm bdy}/c = 0.1$. There is a phase transition from the connected channel to disconnected one, and the capacity drops down by $2S_{\rm bdy} = 0.2c$ at the transition time.
    }
    \label{fig:MMstatic}
\end{figure}

For a static mirror $z(t) = 0$ we have
\begin{align}\label{eq:statictraj}
    p(u) = u\ .
\end{align}
The energy flux \eqref{energy_flux} vanishes, so there are no Hawking radiations in this model.
Instead we expand the interval $R = [0.1, 0.5 + t]$ in time as in figure \ref{fig:static mirror traj general CFT}, which mimics the Rindler observer who feels thermal radiation. 
In this case, there is a phase transition from the connected channel to the disconnected one as shown in the left panel of figure \ref{fig:MMstatic}.
The capacity of entanglement shows a discontinuity at the transition point $t_{\rm phase}$ as in the right panel of figure \ref{fig:MMstatic}.
In contrast to the previous case for an eternal black hole, the discontinuity is negative:
\begin{align}
    \begin{aligned}
    C^{\rm dis}-C^{\rm con}|_{t_{\rm phase}} = -2S_{\rm bdy}\ ,
    \end{aligned}
\end{align}
as the direction of the phase transition is opposite to the previous case.

\section{Discussion}\label{ss:Discussion}

In this paper, we found that the capacity of entanglement 
has a peak or discontinuity at the Page time in two simple models of evaporating black holes, the EOW brane model \cite{Penington:2019kki} (section \ref{ss:ToyModel}) and the moving mirror model \cite{Akal:2020twv} (section \ref{ss:MovingMirror}). 
These observations lead us to conclude that the capacity can be a useful probe of the Hawking radiation in evaporating black holes.
Comparison with other quantum information measures which can capture similar signals, e.g., the reflected entropy or the entanglement wedge cross section \cite{Chandrasekaran:2020qtn,Li:2020ceg} and the relative entropy in charged states \cite{Chen:2020ojn},\footnote{For the topic discussed in \cite{Chen:2020ojn} or a global symmetry violation in the Hawking radiation, see also \cite{Hsin:2020mfa,Harlow:2020bee,Yonekura:2020ino}.} would also be interesting to examine the similarity or difference between the capacity and the other measures.

The simple model calculations elaborated in section \ref{ss:ToyModel} revealed the behavior of the capacity highly depends on the choice of ensembles while the entropy takes almost the same form in both cases.
In the microcanonical ensemble, the capacity has a peak at the Page time ($\log k = \mathbf{S}$) and decays rapidly to zero in the late time.
On the other hand, we find a similar peak at the Page time ($\log k = S_0$) in the canonical ensemble, but the capacity does not decay and approaches to a constant value fixed by the temperature $\beta$ and the brane tension $\mu$ at late time.
It would be worthwhile to further study the dependence of the capacity on the choice of ensembles by introducing chemical potentials to the two-dimensional dilaton gravity.\footnote{For the grand canonical ensembles associated to asymptotic boundaries or baby universes, see \cite{Marolf:2020xie,Arefeva:2019buu}, and to defects in the bulk see \cite{Mertens:2019tcm,Maxfield:2020ale}. See also \cite{Hsin:2020mfa} for charged states in the EOW brane model.}

The capacity in the canonical ensemble of the EOW brane model clearly exhibits a peak around the Page time, but there is a bump after the Page time peak when $S_0$ is small as in the left plot of figure \ref{fig:canonical}.
Note that we use an approximation in the plot which may give rise to an $O((\beta\,G_N)^0)$ error to the entropy  around the Page time in comparing with the exact result \cite{Penington:2019kki}. 
The capacity should also suffer from a similar type of error around the Page time and the exact numerical calculation will reveal whether the bump is caused by an error or not.
If it would not originate from an error, but still remain, the bump could be a signature of a more refined phase transition in the radiation process \cite{Penington:2019kki}.\footnote{
Random pure states in a Floquet model describe thermal process similar to the Hawking radiation, which have prethermalized and chaos phases in addition to the early and late time phases \cite{Chen:2017yzn}.
The peak and bump in the capacity might be associated with a development of these phases.
}

The phase transition between the replica wormholes are caused by non-perturbative instanton corrections to the replica partition functions. 
We believe the capacity of entanglement is a invaluable probe not only for the topology change of replica wormholes but also for other non-perturbative phenomena in non-gravitating theories where replica instanton effect does matter, e.g., confinement in a simple gauge theory with an axion-like coupling to a scalar field  \cite{Ohmori:2021fms,Yonekura:2020ino}.
Other cases where instanton corrections matter are the fixed-area states in AdS/CFT \cite{Dong:2018seb,Akers:2018fow,Marolf:2020vsi} and the random tensor networks \cite{Pastawski:2015qua} (see Appendix E of \cite{Penington:2019kki}). 
A similar calculation as in \cite{Penington:2019kki} may be applied to the capacity of entanglement while it will require more effort due to multiple instanton corrections than the entanglement entropy which only has a one instanton correction.

In the moving mirror models, the differences of the capacities of entanglement between before and after the phase transitions took negative values, except for the case for static mirror probed by an expanding interval.
The sign of the discontinuity is positive when the disconnected channel is favored at early time and taken over to the connected channel at late time, while it is negative when the transition goes in the other way.
It is not clear whether the sign of the discontinuity is related to some nature of the Hawking radiations such as the stability of evaporating black holes.
It would deserve further studies to see if the sign can be constrained by some physical conditions such as unitarity and causality in a broader class of models than those in this paper. 

We have been focused on the exploration of the capacity only in the simple toy models of black holes, but it should be feasible to extend our work to more realistic black holes where we believe the capacity will exhibit a peak or discontinuity when there happens a phase transition between replica wormhole geometries.
We hope to report further investigations of the capacity in a $2d$ CFT coupled to the JT gravity in future work \cite{Kawabata:2021vyo}.

\acknowledgments
We are grateful to K.\,Goto, T.\,Ugajin, T.\,Takayanagi and K.\,Tamaoka for valuable discussions.
The work of T.\,N. was supported in part by the JSPS Grant-in-Aid for Scientific Research (C) No.19K03863 and the JSPS Grant-in-Aid for Scientific Research (A) No.16H02182. 
The work of K.\,W. was supported
by the Grant-in-Aid for JSPS Fellows No.18J00322 and U.S. Department of Energy grant DE-SC0019480 under the HEP-QIS QuantISED program, and by funds from the University of California. The works of K.\,K. and Y.\,O. were supported by Forefront Physics and Mathematics Program to Drive Transformation (FoPM), a World-leading Innovative Graduate Study (WINGS) Program, the University of Tokyo.

\bibliographystyle{JHEP}
\bibliography{ReplicaCapacity}

\providecommand{\href}[2]{#2}\begingroup\raggedright\begin{thebibliography}{100}

\bibitem{Bekenstein:1972tm}
J.~Bekenstein, \emph{{Black holes and the second law}},
  \href{https://doi.org/10.1007/BF02757029}{\emph{Lett. Nuovo Cim.} {\bfseries
  4} (1972) 737--740}.

\bibitem{Bekenstein:1973ur}
J.~D. Bekenstein, \emph{{Black holes and entropy}},
  \href{https://doi.org/10.1103/PhysRevD.7.2333}{\emph{Phys. Rev. D} {\bfseries
  7} (1973) 2333--2346}.

\bibitem{Hawking:1974sw}
S.~Hawking, \emph{{Particle Creation by Black Holes}},
  \href{https://doi.org/10.1007/BF02345020}{\emph{Commun. Math. Phys.}
  {\bfseries 43} (1975) 199--220}.

\bibitem{PhysRevD.14.2460}
S.~W. Hawking, \emph{Breakdown of predictability in gravitational collapse},
  {\emph{Phys. Rev. D} {\bfseries 14} (Nov, 1976) 2460--2473}.

\bibitem{Page:1993wv}
D.~N. Page, \emph{{Information in black hole radiation}},
  \href{https://doi.org/10.1103/PhysRevLett.71.3743}{\emph{Phys. Rev. Lett.}
  {\bfseries 71} (1993) 3743--3746},
  [\href{https://arxiv.org/abs/hep-th/9306083}{{\ttfamily hep-th/9306083}}].

\bibitem{Raju:2020smc}
S.~Raju, \emph{{Lessons from the Information Paradox}},
  \href{https://arxiv.org/abs/2012.05770}{{\ttfamily 2012.05770}}.

\bibitem{Ryu:2006bv}
S.~Ryu and T.~Takayanagi, \emph{{Holographic derivation of entanglement entropy
  from AdS/CFT}},
  \href{https://doi.org/10.1103/PhysRevLett.96.181602}{\emph{Phys. Rev. Lett.}
  {\bfseries 96} (2006) 181602},
  [\href{https://arxiv.org/abs/hep-th/0603001}{{\ttfamily hep-th/0603001}}].

\bibitem{Ryu:2006ef}
S.~Ryu and T.~Takayanagi, \emph{{Aspects of Holographic Entanglement Entropy}},
  \href{https://doi.org/10.1088/1126-6708/2006/08/045}{\emph{JHEP} {\bfseries
  08} (2006) 045}, [\href{https://arxiv.org/abs/hep-th/0605073}{{\ttfamily
  hep-th/0605073}}].

\bibitem{Lewkowycz:2013nqa}
A.~Lewkowycz and J.~Maldacena, \emph{{Generalized gravitational entropy}},
  \href{https://doi.org/10.1007/JHEP08(2013)090}{\emph{JHEP} {\bfseries 08}
  (2013) 090}, [\href{https://arxiv.org/abs/1304.4926}{{\ttfamily 1304.4926}}].

\bibitem{Hubeny:2007xt}
V.~E. Hubeny, M.~Rangamani and T.~Takayanagi, \emph{{A Covariant holographic
  entanglement entropy proposal}},
  \href{https://doi.org/10.1088/1126-6708/2007/07/062}{\emph{JHEP} {\bfseries
  07} (2007) 062}, [\href{https://arxiv.org/abs/0705.0016}{{\ttfamily
  0705.0016}}].

\bibitem{Dong:2016hjy}
X.~Dong, A.~Lewkowycz and M.~Rangamani, \emph{{Deriving covariant holographic
  entanglement}}, \href{https://doi.org/10.1007/JHEP11(2016)028}{\emph{JHEP}
  {\bfseries 11} (2016) 028},
  [\href{https://arxiv.org/abs/1607.07506}{{\ttfamily 1607.07506}}].

\bibitem{Faulkner:2013ana}
T.~Faulkner, A.~Lewkowycz and J.~Maldacena, \emph{{Quantum corrections to
  holographic entanglement entropy}},
  \href{https://doi.org/10.1007/JHEP11(2013)074}{\emph{JHEP} {\bfseries 11}
  (2013) 074}, [\href{https://arxiv.org/abs/1307.2892}{{\ttfamily 1307.2892}}].

\bibitem{Dong:2016fnf}
X.~Dong, \emph{{The Gravity Dual of R\'enyi Entropy}},
  \href{https://doi.org/10.1038/ncomms12472}{\emph{Nature Commun.} {\bfseries
  7} (2016) 12472}, [\href{https://arxiv.org/abs/1601.06788}{{\ttfamily
  1601.06788}}].

\bibitem{Engelhardt:2014gca}
N.~Engelhardt and A.~C. Wall, \emph{{Quantum Extremal Surfaces: Holographic
  Entanglement Entropy beyond the Classical Regime}},
  \href{https://doi.org/10.1007/JHEP01(2015)073}{\emph{JHEP} {\bfseries 01}
  (2015) 073}, [\href{https://arxiv.org/abs/1408.3203}{{\ttfamily 1408.3203}}].

\bibitem{Bekenstein:1974ax}
J.~D. Bekenstein, \emph{{Generalized second law of thermodynamics in black hole
  physics}}, \href{https://doi.org/10.1103/PhysRevD.9.3292}{\emph{Phys. Rev. D}
  {\bfseries 9} (1974) 3292--3300}.

\bibitem{Almheiri:2019hni}
A.~Almheiri, R.~Mahajan, J.~Maldacena and Y.~Zhao, \emph{{The Page curve of
  Hawking radiation from semiclassical geometry}},
  \href{https://doi.org/10.1007/JHEP03(2020)149}{\emph{JHEP} {\bfseries 03}
  (2020) 149}, [\href{https://arxiv.org/abs/1908.10996}{{\ttfamily
  1908.10996}}].

\bibitem{Almheiri:2019qdq}
A.~Almheiri, T.~Hartman, J.~Maldacena, E.~Shaghoulian and A.~Tajdini,
  \emph{{Replica Wormholes and the Entropy of Hawking Radiation}},
  \href{https://doi.org/10.1007/JHEP05(2020)013}{\emph{JHEP} {\bfseries 05}
  (2020) 013}, [\href{https://arxiv.org/abs/1911.12333}{{\ttfamily
  1911.12333}}].

\bibitem{Penington:2019kki}
G.~Penington, S.~H. Shenker, D.~Stanford and Z.~Yang, \emph{{Replica wormholes
  and the black hole interior}},
  \href{https://arxiv.org/abs/1911.11977}{{\ttfamily 1911.11977}}.

\bibitem{Almheiri:2020cfm}
A.~Almheiri, T.~Hartman, J.~Maldacena, E.~Shaghoulian and A.~Tajdini,
  \emph{{The entropy of Hawking radiation}},
  \href{https://arxiv.org/abs/2006.06872}{{\ttfamily 2006.06872}}.

\bibitem{Hartman:2020swn}
T.~Hartman, E.~Shaghoulian and A.~Strominger, \emph{{Islands in Asymptotically
  Flat 2D Gravity}}, \href{https://doi.org/10.1007/JHEP07(2020)022}{\emph{JHEP}
  {\bfseries 07} (2020) 022},
  [\href{https://arxiv.org/abs/2004.13857}{{\ttfamily 2004.13857}}].

\bibitem{almheiri2019islands}
A.~Almheiri, R.~Mahajan and J.~Maldacena, \emph{Islands outside the horizon},
  \href{https://arxiv.org/abs/1910.11077}{{\ttfamily 1910.11077}}.

\bibitem{Balasubramanian:2020xqf}
V.~Balasubramanian, A.~Kar and T.~Ugajin, \emph{{Islands in de Sitter space}},
  \href{https://arxiv.org/abs/2008.05275}{{\ttfamily 2008.05275}}.

\bibitem{hartman2020islands}
T.~Hartman, Y.~Jiang and E.~Shaghoulian, \emph{Islands in cosmology},
  \href{https://arxiv.org/abs/2008.01022}{{\ttfamily 2008.01022}}.

\bibitem{Anegawa:2020ezn}
T.~Anegawa and N.~Iizuka, \emph{{Notes on islands in asymptotically flat 2d
  dilaton black holes}},
  \href{https://doi.org/10.1007/JHEP07(2020)036}{\emph{JHEP} {\bfseries 07}
  (2020) 036}, [\href{https://arxiv.org/abs/2004.01601}{{\ttfamily
  2004.01601}}].

\bibitem{Hashimoto:2020cas}
K.~Hashimoto, N.~Iizuka and Y.~Matsuo, \emph{{Islands in Schwarzschild black
  holes}}, \href{https://doi.org/10.1007/JHEP06(2020)085}{\emph{JHEP}
  {\bfseries 06} (2020) 085},
  [\href{https://arxiv.org/abs/2004.05863}{{\ttfamily 2004.05863}}].

\bibitem{Gautason:2020tmk}
F.~F. Gautason, L.~Schneiderbauer, W.~Sybesma and L.~Thorlacius, \emph{{Page
  Curve for an Evaporating Black Hole}},
  \href{https://doi.org/10.1007/JHEP05(2020)091}{\emph{JHEP} {\bfseries 05}
  (2020) 091}, [\href{https://arxiv.org/abs/2004.00598}{{\ttfamily
  2004.00598}}].

\bibitem{Almheiri:2019psy}
A.~Almheiri, R.~Mahajan and J.~E. Santos, \emph{{Entanglement islands in higher
  dimensions}},
  \href{https://doi.org/10.21468/SciPostPhys.9.1.001}{\emph{SciPost Phys.}
  {\bfseries 9} (2020) 001},
  [\href{https://arxiv.org/abs/1911.09666}{{\ttfamily 1911.09666}}].

\bibitem{Alishahiha:2020qza}
M.~Alishahiha, A.~Faraji~Astaneh and A.~Naseh, \emph{{Island in the Presence of
  Higher Derivative Terms}},
  \href{https://arxiv.org/abs/2005.08715}{{\ttfamily 2005.08715}}.

\bibitem{Ling:2020laa}
Y.~Ling, Y.~Liu and Z.-Y. Xian, \emph{{Island in Charged Black Holes}},
  \href{https://arxiv.org/abs/2010.00037}{{\ttfamily 2010.00037}}.

\bibitem{Bhattacharya:2020uun}
A.~Bhattacharya, A.~Chanda, S.~Maulik, C.~Northe and S.~Roy, \emph{{Topological
  shadows and complexity of islands in multiboundary wormholes}},
  \href{https://arxiv.org/abs/2010.04134}{{\ttfamily 2010.04134}}.

\bibitem{Akers:2019nfi}
C.~Akers, N.~Engelhardt and D.~Harlow, \emph{{Simple holographic models of
  black hole evaporation}},
  \href{https://doi.org/10.1007/JHEP08(2020)032}{\emph{JHEP} {\bfseries 08}
  (2020) 032}, [\href{https://arxiv.org/abs/1910.00972}{{\ttfamily
  1910.00972}}].

\bibitem{Chen:2019uhq}
H.~Z. Chen, Z.~Fisher, J.~Hernandez, R.~C. Myers and S.-M. Ruan,
  \emph{{Information Flow in Black Hole Evaporation}},
  \href{https://doi.org/10.1007/JHEP03(2020)152}{\emph{JHEP} {\bfseries 03}
  (2020) 152}, [\href{https://arxiv.org/abs/1911.03402}{{\ttfamily
  1911.03402}}].

\bibitem{Nomura:2019qps}
Y.~Nomura, \emph{{Spacetime and Universal Soft Modes --- Black Holes and
  Beyond}}, \href{https://doi.org/10.1103/PhysRevD.101.066024}{\emph{Phys. Rev.
  D} {\bfseries 101} (2020) 066024},
  [\href{https://arxiv.org/abs/1908.05728}{{\ttfamily 1908.05728}}].

\bibitem{Suzuki:2019xdq}
Y.~Suzuki, T.~Takayanagi and K.~Umemoto, \emph{{Entanglement Wedges from the
  Information Metric in Conformal Field Theories}},
  \href{https://doi.org/10.1103/PhysRevLett.123.221601}{\emph{Phys. Rev. Lett.}
  {\bfseries 123} (2019) 221601},
  [\href{https://arxiv.org/abs/1908.09939}{{\ttfamily 1908.09939}}].

\bibitem{Kusuki:2019hcg}
Y.~Kusuki, Y.~Suzuki, T.~Takayanagi and K.~Umemoto, \emph{{Looking at Shadows
  of Entanglement Wedges}},
  \href{https://doi.org/10.1093/ptep/ptaa152}{\emph{PTEP} {\bfseries 2020}
  (2020) 11B105}, [\href{https://arxiv.org/abs/1912.08423}{{\ttfamily
  1912.08423}}].

\bibitem{Rozali:2019day}
M.~Rozali, J.~Sully, M.~Van~Raamsdonk, C.~Waddell and D.~Wakeham,
  \emph{{Information radiation in BCFT models of black holes}},
  \href{https://doi.org/10.1007/JHEP05(2020)004}{\emph{JHEP} {\bfseries 05}
  (2020) 004}, [\href{https://arxiv.org/abs/1910.12836}{{\ttfamily
  1910.12836}}].

\bibitem{Balasubramanian:2020coy}
V.~Balasubramanian, A.~Kar and T.~Ugajin, \emph{{Entanglement between two
  disjoint universes}},  \href{https://arxiv.org/abs/2008.05274}{{\ttfamily
  2008.05274}}.

\bibitem{Chen:2020jvn}
H.~Z. Chen, Z.~Fisher, J.~Hernandez, R.~C. Myers and S.-M. Ruan,
  \emph{{Evaporating Black Holes Coupled to a Thermal Bath}},
  \href{https://doi.org/10.1007/JHEP01(2021)065}{\emph{JHEP} {\bfseries 01}
  (2021) 065}, [\href{https://arxiv.org/abs/2007.11658}{{\ttfamily
  2007.11658}}].

\bibitem{Bak:2020enw}
D.~Bak, C.~Kim, S.-H. Yi and J.~Yoon, \emph{{Unitarity of Entanglement and
  Islands in Two-Sided Janus Black Holes}},
  \href{https://arxiv.org/abs/2006.11717}{{\ttfamily 2006.11717}}.

\bibitem{Agon:2020fqs}
C.~A. Ag\'on, S.~F. Lokhande and J.~F. Pedraza, \emph{{Local quenches, bulk
  entanglement entropy and a unitary Page curve}},
  \href{https://doi.org/10.1007/JHEP08(2020)152}{\emph{JHEP} {\bfseries 08}
  (2020) 152}, [\href{https://arxiv.org/abs/2004.15010}{{\ttfamily
  2004.15010}}].

\bibitem{Balasubramanian:2020hfs}
V.~Balasubramanian, A.~Kar, O.~Parrikar, G.~S\'arosi and T.~Ugajin,
  \emph{{Geometric secret sharing in a model of Hawking radiation}},
  \href{https://arxiv.org/abs/2003.05448}{{\ttfamily 2003.05448}}.

\bibitem{Krishnan:2020oun}
C.~Krishnan, V.~Patil and J.~Pereira, \emph{{Page Curve and the Information
  Paradox in Flat Space}},  \href{https://arxiv.org/abs/2005.02993}{{\ttfamily
  2005.02993}}.

\bibitem{Krishnan:2020fer}
C.~Krishnan, \emph{{Critical Islands}},
  \href{https://doi.org/10.1007/JHEP01(2021)179}{\emph{JHEP} {\bfseries 01}
  (2021) 179}, [\href{https://arxiv.org/abs/2007.06551}{{\ttfamily
  2007.06551}}].

\bibitem{Li:2020ceg}
T.~Li, J.~Chu and Y.~Zhou, \emph{{Reflected Entropy for an Evaporating Black
  Hole}},  \href{https://arxiv.org/abs/2006.10846}{{\ttfamily 2006.10846}}.

\bibitem{Chandrasekaran:2020qtn}
V.~Chandrasekaran, M.~Miyaji and P.~Rath, \emph{{Including contributions from
  entanglement islands to the reflected entropy}},
  \href{https://doi.org/10.1103/PhysRevD.102.086009}{\emph{Phys. Rev. D}
  {\bfseries 102} (2020) 086009},
  [\href{https://arxiv.org/abs/2006.10754}{{\ttfamily 2006.10754}}].

\bibitem{Dong:2020uxp}
X.~Dong, X.-L. Qi, Z.~Shangnan and Z.~Yang, \emph{{Effective entropy of quantum
  fields coupled with gravity}},
  \href{https://doi.org/10.1007/JHEP10(2020)052}{\emph{JHEP} {\bfseries 10}
  (2020) 052}, [\href{https://arxiv.org/abs/2007.02987}{{\ttfamily
  2007.02987}}].

\bibitem{Geng:2020qvw}
H.~Geng and A.~Karch, \emph{{Massive islands}},
  \href{https://doi.org/10.1007/JHEP09(2020)121}{\emph{JHEP} {\bfseries 09}
  (2020) 121}, [\href{https://arxiv.org/abs/2006.02438}{{\ttfamily
  2006.02438}}].

\bibitem{Chen:2020uac}
H.~Z. Chen, R.~C. Myers, D.~Neuenfeld, I.~A. Reyes and J.~Sandor,
  \emph{{Quantum Extremal Islands Made Easy, Part I: Entanglement on the
  Brane}}, \href{https://doi.org/10.1007/JHEP10(2020)166}{\emph{JHEP}
  {\bfseries 10} (2020) 166},
  [\href{https://arxiv.org/abs/2006.04851}{{\ttfamily 2006.04851}}].

\bibitem{Chen:2020hmv}
H.~Z. Chen, R.~C. Myers, D.~Neuenfeld, I.~A. Reyes and J.~Sandor,
  \emph{{Quantum Extremal Islands Made Easy, Part II: Black Holes on the
  Brane}}, \href{https://doi.org/10.1007/JHEP12(2020)025}{\emph{JHEP}
  {\bfseries 12} (2020) 025},
  [\href{https://arxiv.org/abs/2010.00018}{{\ttfamily 2010.00018}}].

\bibitem{Hernandez:2020nem}
J.~Hernandez, R.~C. Myers and S.-M. Ruan, \emph{{Quantum Extremal Islands Made
  Easy, PartIII: Complexity on the Brane}},
  \href{https://arxiv.org/abs/2010.16398}{{\ttfamily 2010.16398}}.

\bibitem{Akal:2020ujg}
I.~Akal, \emph{{Universality, intertwiners and black hole information}},
  \href{https://arxiv.org/abs/2010.12565}{{\ttfamily 2010.12565}}.

\bibitem{Kirklin:2020zic}
J.~Kirklin, \emph{{Islands and Uhlmann phase: Explicit recovery of classical
  information from evaporating black holes}},
  \href{https://arxiv.org/abs/2011.07086}{{\ttfamily 2011.07086}}.

\bibitem{Liu:2020gnp}
H.~Liu and S.~Vardhan, \emph{{A dynamical mechanism for the Page curve from
  quantum chaos}},  \href{https://arxiv.org/abs/2002.05734}{{\ttfamily
  2002.05734}}.

\bibitem{Piroli:2020dlx}
L.~Piroli, C.~S\"underhauf and X.-L. Qi, \emph{{A Random Unitary Circuit Model
  for Black Hole Evaporation}},
  \href{https://doi.org/10.1007/JHEP04(2020)063}{\emph{JHEP} {\bfseries 04}
  (2020) 063}, [\href{https://arxiv.org/abs/2002.09236}{{\ttfamily
  2002.09236}}].

\bibitem{Nomura:2020ska}
Y.~Nomura, \emph{{Black Hole Interior in Unitary Gauge Construction}},
  \href{https://arxiv.org/abs/2010.15827}{{\ttfamily 2010.15827}}.

\bibitem{Pollack:2020gfa}
J.~Pollack, M.~Rozali, J.~Sully and D.~Wakeham, \emph{{Eigenstate
  Thermalization and Disorder Averaging in Gravity}},
  \href{https://doi.org/10.1103/PhysRevLett.125.021601}{\emph{Phys. Rev. Lett.}
  {\bfseries 125} (2020) 021601},
  [\href{https://arxiv.org/abs/2002.02971}{{\ttfamily 2002.02971}}].

\bibitem{Marolf:2020xie}
D.~Marolf and H.~Maxfield, \emph{{Transcending the ensemble: baby universes,
  spacetime wormholes, and the order and disorder of black hole information}},
  \href{https://doi.org/10.1007/JHEP08(2020)044}{\emph{JHEP} {\bfseries 08}
  (2020) 044}, [\href{https://arxiv.org/abs/2002.08950}{{\ttfamily
  2002.08950}}].

\bibitem{Chakravarty:2020wdm}
J.~Chakravarty, \emph{{Overcounting of interior excitations: A resolution to
  the bags of gold paradox in AdS}},
  \href{https://arxiv.org/abs/2010.03575}{{\ttfamily 2010.03575}}.

\bibitem{Marolf:2020rpm}
D.~Marolf and H.~Maxfield, \emph{{Observations of Hawking radiation: the Page
  curve and baby universes}},
  \href{https://arxiv.org/abs/2010.06602}{{\ttfamily 2010.06602}}.

\bibitem{Bousso:2020kmy}
R.~Bousso and E.~Wildenhain, \emph{{Gravity/ensemble duality}},
  \href{https://doi.org/10.1103/PhysRevD.102.066005}{\emph{Phys. Rev. D}
  {\bfseries 102} (2020) 066005},
  [\href{https://arxiv.org/abs/2006.16289}{{\ttfamily 2006.16289}}].

\bibitem{Stanford:2020wkf}
D.~Stanford, \emph{{More quantum noise from wormholes}},
  \href{https://arxiv.org/abs/2008.08570}{{\ttfamily 2008.08570}}.

\bibitem{Giddings:2020yes}
S.~B. Giddings and G.~J. Turiaci, \emph{{Wormhole calculus, replicas, and
  entropies}}, \href{https://doi.org/10.1007/JHEP09(2020)194}{\emph{JHEP}
  {\bfseries 09} (2020) 194},
  [\href{https://arxiv.org/abs/2004.02900}{{\ttfamily 2004.02900}}].

\bibitem{Chen:2020tes}
Y.~Chen, V.~Gorbenko and J.~Maldacena, \emph{{Bra-ket wormholes in
  gravitationally prepared states}},
  \href{https://arxiv.org/abs/2007.16091}{{\ttfamily 2007.16091}}.

\bibitem{Karlsson:2020uga}
A.~Karlsson, \emph{{Replica wormhole and island incompatibility with monogamy
  of entanglement}},  \href{https://arxiv.org/abs/2007.10523}{{\ttfamily
  2007.10523}}.

\bibitem{Engelhardt:2020qpv}
N.~Engelhardt, S.~Fischetti and A.~Maloney, \emph{{Free Energy from Replica
  Wormholes}},  \href{https://arxiv.org/abs/2007.07444}{{\ttfamily
  2007.07444}}.

\bibitem{Verlinde:2020upt}
H.~Verlinde, \emph{{ER = EPR revisited: On the Entropy of an Einstein-Rosen
  Bridge}},  \href{https://arxiv.org/abs/2003.13117}{{\ttfamily 2003.13117}}.

\bibitem{Harlow:2020bee}
D.~Harlow and E.~Shaghoulian, \emph{{Global symmetry, Euclidean gravity, and
  the black hole information problem}},
  \href{https://arxiv.org/abs/2010.10539}{{\ttfamily 2010.10539}}.

\bibitem{Chen:2020ojn}
Y.~Chen and H.~W. Lin, \emph{{Signatures of global symmetry violation in
  relative entropies and replica wormholes}},
  \href{https://arxiv.org/abs/2011.06005}{{\ttfamily 2011.06005}}.

\bibitem{Hsin:2020mfa}
P.-S. Hsin, L.~V. Iliesiu and Z.~Yang, \emph{{A violation of global symmetries
  from replica wormholes and the fate of black hole remnants}},
  \href{https://arxiv.org/abs/2011.09444}{{\ttfamily 2011.09444}}.

\bibitem{Akal:2020wfl}
I.~Akal, Y.~Kusuki, T.~Takayanagi and Z.~Wei, \emph{{Codimension two holography
  for wedges}}, \href{https://doi.org/10.1103/PhysRevD.102.126007}{\emph{Phys.
  Rev. D} {\bfseries 102} (2020) 126007},
  [\href{https://arxiv.org/abs/2007.06800}{{\ttfamily 2007.06800}}].

\bibitem{Murdia:2020iac}
C.~Murdia, Y.~Nomura and P.~Rath, \emph{{Coarse-Graining Holographic States: A
  Semiclassical Flow in General Spacetimes}},
  \href{https://doi.org/10.1103/PhysRevD.102.086001}{\emph{Phys. Rev. D}
  {\bfseries 102} (2020) 086001},
  [\href{https://arxiv.org/abs/2008.01755}{{\ttfamily 2008.01755}}].

\bibitem{Goto:2020wnk}
K.~Goto, T.~Hartman and A.~Tajdini, \emph{{Replica wormholes for an evaporating
  2D black hole}},  \href{https://arxiv.org/abs/2011.09043}{{\ttfamily
  2011.09043}}.

\bibitem{Geng:2020fxl}
H.~Geng, A.~Karch, C.~Perez-Pardavila, S.~Raju, L.~Randall, M.~Riojas et~al.,
  \emph{{Information Transfer with a Gravitating Bath}},
  \href{https://arxiv.org/abs/2012.04671}{{\ttfamily 2012.04671}}.

\bibitem{Caceres:2020jcn}
E.~Caceres, A.~Kundu, A.~K. Patra and S.~Shashi, \emph{{Warped Information and
  Entanglement Islands in AdS/WCFT}},
  \href{https://arxiv.org/abs/2012.05425}{{\ttfamily 2012.05425}}.

\bibitem{Manu:2020tty}
A.~Manu, K.~Narayan and P.~Paul, \emph{{Cosmological singularities,
  entanglement and quantum extremal surfaces}},
  \href{https://doi.org/10.1007/JHEP04(2021)200}{\emph{JHEP} {\bfseries 04}
  (2021) 200}, [\href{https://arxiv.org/abs/2012.07351}{{\ttfamily
  2012.07351}}].

\bibitem{Miao:2020oey}
R.-X. Miao, \emph{{An Exact Construction of Codimension two Holography}},
  \href{https://doi.org/10.1007/JHEP01(2021)150}{\emph{JHEP} {\bfseries 01}
  (2021) 150}, [\href{https://arxiv.org/abs/2009.06263}{{\ttfamily
  2009.06263}}].

\bibitem{Miao:2021ual}
R.-X. Miao, \emph{{Codimension-n Holography for the Cones}},
  \href{https://arxiv.org/abs/2101.10031}{{\ttfamily 2101.10031}}.

\bibitem{Yao:2010woi}
H.~Yao and X.-L. Qi, \emph{{Entanglement entropy and entanglement spectrum of
  the Kitaev model}},
  \href{https://doi.org/10.1103/PhysRevLett.105.080501}{\emph{Phys. Rev. Lett.}
  {\bfseries 105} (2010) 080501},
  [\href{https://arxiv.org/abs/1001.1165}{{\ttfamily 1001.1165}}].

\bibitem{Nakaguchi:2016zqi}
Y.~Nakaguchi and T.~Nishioka, \emph{{A holographic proof of R\'enyi entropic
  inequalities}}, \href{https://doi.org/10.1007/JHEP12(2016)129}{\emph{JHEP}
  {\bfseries 12} (2016) 129},
  [\href{https://arxiv.org/abs/1606.08443}{{\ttfamily 1606.08443}}].

\bibitem{deBoer:2018mzv}
J.~De~Boer, J.~J\"arvel\"a and E.~Keski-Vakkuri, \emph{{Aspects of capacity of
  entanglement}}, \href{https://doi.org/10.1103/PhysRevD.99.066012}{\emph{Phys.
  Rev. D} {\bfseries 99} (2019) 066012},
  [\href{https://arxiv.org/abs/1807.07357}{{\ttfamily 1807.07357}}].

\bibitem{Nakagawa:2017wis}
Y.~O. Nakagawa and S.~Furukawa, \emph{{Capacity of entanglement and the
  distribution of density matrix eigenvalues in gapless systems}},
  \href{https://doi.org/10.1103/PhysRevB.96.205108}{\emph{Phys. Rev. B}
  {\bfseries 96} (2017) 205108},
  [\href{https://arxiv.org/abs/1708.08924}{{\ttfamily 1708.08924}}].

\bibitem{Akal:2020twv}
I.~Akal, Y.~Kusuki, N.~Shiba, T.~Takayanagi and Z.~Wei, \emph{{Entanglement
  entropy in holographic moving mirror and Page curve}},
  \href{https://arxiv.org/abs/2011.12005}{{\ttfamily 2011.12005}}.

\bibitem{Teitelboim:1983ux}
C.~Teitelboim, \emph{{Gravitation and Hamiltonian Structure in Two Space-Time
  Dimensions}}, \href{https://doi.org/10.1016/0370-2693(83)90012-6}{\emph{Phys.
  Lett. B} {\bfseries 126} (1983) 41--45}.

\bibitem{Jackiw:1984je}
R.~Jackiw, \emph{{Lower Dimensional Gravity}},
  \href{https://doi.org/10.1016/0550-3213(85)90448-1}{\emph{Nucl. Phys. B}
  {\bfseries 252} (1985) 343--356}.

\bibitem{Kourkoulou:2017zaj}
I.~Kourkoulou and J.~Maldacena, \emph{{Pure states in the SYK model and
  nearly-$AdS_2$ gravity}},  \href{https://arxiv.org/abs/1707.02325}{{\ttfamily
  1707.02325}}.

\bibitem{Page:1993df}
D.~N. Page, \emph{{Average entropy of a subsystem}},
  \href{https://doi.org/10.1103/PhysRevLett.71.1291}{\emph{Phys. Rev. Lett.}
  {\bfseries 71} (1993) 1291--1294},
  [\href{https://arxiv.org/abs/gr-qc/9305007}{{\ttfamily gr-qc/9305007}}].

\bibitem{Davies:1976hi}
P.~C.~W. Davies and S.~A. Fulling, \emph{{Radiation from a moving mirror in
  two-dimensional space-time conformal anomaly}}, {\emph{Proc. Roy. Soc. Lond.
  A} {\bfseries 348} (1976) 393--414}.

\bibitem{birrell1984quantum}
N.~D. Birrell and P.~C.~W. Davies, \emph{Quantum fields in curved space}.
\newblock No.~7. Cambridge university press, 1984.

\bibitem{Carlitz:1986ng}
R.~D. Carlitz and R.~S. Willey, \emph{{The Lifetime of a Black Hole}},
  \href{https://doi.org/10.1103/PhysRevD.36.2336}{\emph{Phys. Rev. D}
  {\bfseries 36} (1987) 2336}.

\bibitem{Carlitz:1986nh}
R.~D. Carlitz and R.~S. Willey, \emph{{Reflections on moving mirrors}},
  \href{https://doi.org/10.1103/PhysRevD.36.2327}{\emph{Phys. Rev. D}
  {\bfseries 36} (1987) 2327--2335}.

\bibitem{Wilczek:1993jn}
F.~Wilczek, \emph{{Quantum purity at a small price: Easing a black hole
  paradox}},  in \emph{{International Symposium on Black holes, Membranes,
  Wormholes and Superstrings}}, pp.~1--21, 2, 1993,
  \href{https://arxiv.org/abs/hep-th/9302096}{{\ttfamily hep-th/9302096}}.

\bibitem{Bianchi:2014qua}
E.~Bianchi and M.~Smerlak, \emph{{Entanglement entropy and negative energy in
  two dimensions}},
  \href{https://doi.org/10.1103/PhysRevD.90.041904}{\emph{Phys. Rev. D}
  {\bfseries 90} (2014) 041904},
  [\href{https://arxiv.org/abs/1404.0602}{{\ttfamily 1404.0602}}].

\bibitem{Hotta:2015huj}
M.~Hotta and A.~Sugita, \emph{{The Fall of Black Hole Firewall: Natural
  Nonmaximal Entanglement for Page Curve}},
  \href{https://doi.org/10.1093/ptep/ptv170}{\emph{PTEP} {\bfseries 2015}
  (2015) 123B04}, [\href{https://arxiv.org/abs/1505.05870}{{\ttfamily
  1505.05870}}].

\bibitem{Good:2016atu}
M.~R.~R. Good, K.~Yelshibekov and Y.~C. Ong, \emph{{On Horizonless Temperature
  with an Accelerating Mirror}},
  \href{https://doi.org/10.1007/JHEP03(2017)013}{\emph{JHEP} {\bfseries 03}
  (2017) 013}, [\href{https://arxiv.org/abs/1611.00809}{{\ttfamily
  1611.00809}}].

\bibitem{Chen:2017lum}
P.~Chen and D.-h. Yeom, \emph{{Entropy evolution of moving mirrors and the
  information loss problem}},
  \href{https://doi.org/10.1103/PhysRevD.96.025016}{\emph{Phys. Rev. D}
  {\bfseries 96} (2017) 025016},
  [\href{https://arxiv.org/abs/1704.08613}{{\ttfamily 1704.08613}}].

\bibitem{Good:2019tnf}
M.~R.~R. Good, E.~V. Linder and F.~Wilczek, \emph{{Moving mirror model for
  quasithermal radiation fields}},
  \href{https://doi.org/10.1103/PhysRevD.101.025012}{\emph{Phys. Rev. D}
  {\bfseries 101} (2020) 025012},
  [\href{https://arxiv.org/abs/1909.01129}{{\ttfamily 1909.01129}}].

\bibitem{Calabrese:2004eu}
P.~Calabrese and J.~L. Cardy, \emph{{Entanglement entropy and quantum field
  theory}}, \href{https://doi.org/10.1088/1742-5468/2004/06/P06002}{\emph{J.
  Stat. Mech.} {\bfseries 0406} (2004) P06002},
  [\href{https://arxiv.org/abs/hep-th/0405152}{{\ttfamily hep-th/0405152}}].

\bibitem{Affleck:1991tk}
I.~Affleck and A.~W. Ludwig, \emph{{Universal noninteger ``ground state
  degeneracy" in critical quantum systems}},
  \href{https://doi.org/10.1103/PhysRevLett.67.161}{\emph{Phys. Rev. Lett.}
  {\bfseries 67} (1991) 161--164}.

\bibitem{Sully:2020pza}
J.~Sully, M.~Van~Raamsdonk and D.~Wakeham, \emph{{BCFT entanglement entropy at
  large central charge and the black hole interior}},
  \href{https://arxiv.org/abs/2004.13088}{{\ttfamily 2004.13088}}.

\bibitem{Takayanagi:2011zk}
T.~Takayanagi, \emph{{Holographic Dual of BCFT}},
  \href{https://doi.org/10.1103/PhysRevLett.107.101602}{\emph{Phys. Rev. Lett.}
  {\bfseries 107} (2011) 101602},
  [\href{https://arxiv.org/abs/1105.5165}{{\ttfamily 1105.5165}}].

\bibitem{Fujita:2011fp}
M.~Fujita, T.~Takayanagi and E.~Tonni, \emph{{Aspects of AdS/BCFT}},
  \href{https://doi.org/10.1007/JHEP11(2011)043}{\emph{JHEP} {\bfseries 11}
  (2011) 043}, [\href{https://arxiv.org/abs/1108.5152}{{\ttfamily 1108.5152}}].

\bibitem{Yonekura:2020ino}
K.~Yonekura, \emph{{Topological violation of global symmetries in quantum
  gravity}},  \href{https://arxiv.org/abs/2011.11868}{{\ttfamily 2011.11868}}.

\bibitem{Arefeva:2019buu}
I.~Aref'eva and I.~Volovich, \emph{{Gas of Baby Universes in JT Gravity and
  Matrix Models}}, \href{https://doi.org/10.3390/sym12060975}{\emph{Symmetry}
  {\bfseries 12} (2020) 975},
  [\href{https://arxiv.org/abs/1905.08207}{{\ttfamily 1905.08207}}].

\bibitem{Mertens:2019tcm}
T.~G. Mertens and G.~J. Turiaci, \emph{{Defects in Jackiw-Teitelboim Quantum
  Gravity}}, \href{https://doi.org/10.1007/JHEP08(2019)127}{\emph{JHEP}
  {\bfseries 08} (2019) 127},
  [\href{https://arxiv.org/abs/1904.05228}{{\ttfamily 1904.05228}}].

\bibitem{Maxfield:2020ale}
H.~Maxfield and G.~J. Turiaci, \emph{{The path integral of 3D gravity near
  extremality; or, JT gravity with defects as a matrix integral}},
  \href{https://doi.org/10.1007/JHEP01(2021)118}{\emph{JHEP} {\bfseries 01}
  (2021) 118}, [\href{https://arxiv.org/abs/2006.11317}{{\ttfamily
  2006.11317}}].

\bibitem{Chen:2017yzn}
X.~Chen and A.~W.~W. Ludwig, \emph{{Universal Spectral Correlations in the
  Chaotic Wave Function, and the Development of Quantum Chaos}},
  \href{https://doi.org/10.1103/PhysRevB.98.064309}{\emph{Phys. Rev. B}
  {\bfseries 98} (2018) 064309},
  [\href{https://arxiv.org/abs/1710.02686}{{\ttfamily 1710.02686}}].

\bibitem{Ohmori:2021fms}
K.~Ohmori, \emph{{Replica Instantons from Axion-like Coupling}},
  \href{https://arxiv.org/abs/2101.07854}{{\ttfamily 2101.07854}}.

\bibitem{Dong:2018seb}
X.~Dong, D.~Harlow and D.~Marolf, \emph{{Flat entanglement spectra in
  fixed-area states of quantum gravity}},
  \href{https://doi.org/10.1007/JHEP10(2019)240}{\emph{JHEP} {\bfseries 10}
  (2019) 240}, [\href{https://arxiv.org/abs/1811.05382}{{\ttfamily
  1811.05382}}].

\bibitem{Akers:2018fow}
C.~Akers and P.~Rath, \emph{{Holographic R\'enyi Entropy from Quantum Error
  Correction}}, \href{https://doi.org/10.1007/JHEP05(2019)052}{\emph{JHEP}
  {\bfseries 05} (2019) 052},
  [\href{https://arxiv.org/abs/1811.05171}{{\ttfamily 1811.05171}}].

\bibitem{Marolf:2020vsi}
D.~Marolf, S.~Wang and Z.~Wang, \emph{{Probing phase transitions of holographic
  entanglement entropy with fixed area states}},
  \href{https://doi.org/10.1007/JHEP12(2020)084}{\emph{JHEP} {\bfseries 12}
  (2020) 084}, [\href{https://arxiv.org/abs/2006.10089}{{\ttfamily
  2006.10089}}].

\bibitem{Pastawski:2015qua}
F.~Pastawski, B.~Yoshida, D.~Harlow and J.~Preskill, \emph{{Holographic quantum
  error-correcting codes: Toy models for the bulk/boundary correspondence}},
  \href{https://doi.org/10.1007/JHEP06(2015)149}{\emph{JHEP} {\bfseries 06}
  (2015) 149}, [\href{https://arxiv.org/abs/1503.06237}{{\ttfamily
  1503.06237}}].

\bibitem{Kawabata:2021vyo}
K.~Kawabata, T.~Nishioka, Y.~Okuyama and K.~Watanabe, \emph{{Replica wormholes
  and capacity of entanglement}},
  \href{https://arxiv.org/abs/2105.08396}{{\ttfamily 2105.08396}}.

\end{thebibliography}\endgroup

\end{document}